\documentclass[accepted,specialissue]{melba}
\usepackage{amsmath,amssymb,amsfonts}
\usepackage{algorithmic}
\usepackage{graphicx}
\usepackage{subcaption}
\usepackage{url}
\usepackage{booktabs}
\usepackage{placeins}

\usepackage{textcomp}
\def\BibTeX{{\rm B\kern-.05em{\sc i\kern-.025em b}\kern-.08em
    T\kern-.1667em\lower.7ex\hbox{E}\kern-.125emX}}

\melbaid{2026:025}  % This is provided upon by the publishing editor
\doi{10.59275/j.melba.2026-8194}
\melbaauthors{Gao, Yurk, and Abu-Mostafa}  % Note: this one is also used to set the pdf 'authors' metadata
\email{zgao2@caltech.edu}
\volume{2026}
\firstpageno{508}  % Communicated by the publishing editor
\melbayear{2026}  % The publication year
\datesubmitted{2025-12-30}  % Date submitted to MELBA: mm/yyyy
\datepublished{2026-08-27}  % Today's date: mm/yyyy

\melbaspecialissue{Medical Imaging with Deep Learning (MIDL) 2025}
\melbaspecialissueeditors{Lisa Koch, Ronald M. Summers, Chen Chen, Yan Zhuang}

\ShortHeadings{Learning from Scarce Labels: Multi-View Echocardiography for Ejection Fraction Prediction}{Gao, Yurk, and Abu-Mostafa}

\title{Learning from Scarce Labels: Multi-View Echocardiography for Ejection Fraction Prediction}

\author{
	\firstname Zhiyuan \surname Gao\aff{1}\orcid{0009-0005-7633-7149},
	\firstname Dominic \surname Yurk\aff{2}\orcid{0000-0002-2276-4189},
	\firstname Yaser S. \surname Abu-Mostafa\aff{1}
}
\affiliations{% <- trailing '%' to avoid unwanted indent
	\num 1 \addr Electrical Engineering Department, California Institute of Technology, Pasadena, CA 91125, USA \\
	\num 2 \addr Asari AI, San Francisco, CA 94131, USA
}

\abstract{%
 We present, to the best of our knowledge, the first publicly available resource for predicting left ventricular ejection fraction (EF) from parasternal long-axis (PLAX) echocardiography. Because no PLAX–EF datasets previously existed, our work focuses on an~innovative data generation strategy to overcome this scarcity. By leveraging a~time-based correlation between clinical notes and echocardiographic videos, combined with fine-tuning view classifiers and proxy labeling, we created a~labeled dataset of over 25,000 PLAX videos. This enables us to train the first reproducible PLAX EF model, achieving a mean absolute error (MAE) of 6.86\%. Given that apical four-chamber (A4C) methods, the clinical standard, report MAE values of 6\%-7\%, our results demonstrate that EF estimation from PLAX views is both feasible and clinically relevant. This surpasses the performance of existing methods and provides a~clinically relevant solution for situations where apical views may not be feasible. Going further, we demonstrate that combining PLAX and A4C predictions via simple unweighted late fusion improves both single-view baselines to a 6.37\% MAE, underscoring the value of multi-view integration. To promote continued research, we release the dataset labels, trained models, and runnable demos on GitHub, Hugging Face, and Google Colab:  \url{https://github.com/Jeffrey4899/PLAX_EF_Labels_202509}.
}


\begin{document}

% top matter
\twocolumn[\maketitle]
\section{Introduction}
\label{sec:introduction}
\enluminure{C}{ardiovascular} disease remains a leading cause of mortality worldwide \citep{doi:10.1161/CIR.0000000000001123}. Transthoracic echocardiography is a ubiquitous, non-invasive imaging modality for cardiac assessment, and left ventricular ejection fraction (EF) is one of the most clinically actionable summary measures of systolic function. EF informs diagnosis, treatment selection, and longitudinal monitoring across conditions such as heart failure and cardiomyopathies.

In routine clinical reporting, EF is most commonly derived from apical views (notably apical four-chamber, A4C), which provide favorable geometry for standardized volumetric measurements (Section~\ref{sec:background}). However, high-quality apical acquisitions can be difficult in real-world settings, including poor acoustic windows, limited patient positioning, and time-constrained bedside/point-of-care scans. In contrast, the parasternal long-axis (PLAX) view is routinely acquired and often easier to obtain in practice, and clinicians frequently use PLAX for rapid qualitative assessment of LV function \citep{russell2022plaxIsolation}. Despite its ubiquity, there is no widely adopted protocol for reporting EF directly from PLAX cine videos, and consequently there is limited \emph{public} supervision linking PLAX to EF at scale.

This creates a major bottleneck for machine learning: while large public datasets exist for A4C EF regression (e.g., EchoNet-Dynamic \citep{echonet-dynamic}), public PLAX datasets are typically labeled for \emph{other} targets (e.g., LVH-related phenotypes in EchoNet-LVH \citep{echonet-LVH}), and large clinical archives (e.g., MIMIC-IV-Echo) provide videos without view labels \citep{mimic_iv_echo,physionet}. Prior PLAX EF efforts are therefore either proprietary/closed or rely on indirect measurement pipelines; reproducible PLAX EF benchmarks remain scarce (reviewed in Section~\ref{sec:related_work}).

\paragraph{Overview.}
We address the PLAX label scarcity problem by constructing, to the best of our knowledge, the first publicly released resource that associates PLAX cine videos with EF labels at scale using only publicly accessible datasets. Our pipeline combines:
(i) robust video view classification to mine A4C and PLAX from large unlabeled repositories,
(ii) time-based correlation between studies and clinical notes to form an independent, note-derived ground-truth cohort, and
(iii) \emph{proxy supervision} in which A4C-based EF predictions provide study-level labels for PLAX training when direct PLAX EF annotations are unavailable.
Beyond single-view modeling, we demonstrate that simple study-level late fusion of A4C and PLAX predictions yields consistent gains, supporting multi-view EF estimation.

\paragraph{Contributions.}
Concretely, we make the following contributions:
\begin{itemize}
    \item \textbf{A public PLAX--EF resource.} We release label files and end-to-end artifacts that enable reproducible PLAX EF research using MIMIC-IV resources, including instructions for locating the corresponding samples under the data-use agreement \citep{mimic_iv_echo,mimic_iv_note,physionet}.
    \item \textbf{A reproducible PLAX EF benchmark.} Training on large-scale proxy-supervised PLAX data and evaluating on a note-derived ground-truth cohort, our PLAX ensemble achieves \textbf{6.86\% MAE} at the study level.
    \item \textbf{Multi-view EF estimation.} Unweighted late fusion of A4C and PLAX predictions improves both single-view baselines, achieving \textbf{6.37\% MAE} and Pearson correlation 0.709 on studies containing both views.
    \item \textbf{Open artifacts for adoption and extension.} We release runnable demos and pretrained models (GitHub, Hugging Face, and Colab) to support independent validation and downstream research \citep{plax_labels_github}.
\end{itemize}

\paragraph{Extensions over the MIDL 2025 proceedings version.}
As required for MELBA submissions extending a conference paper, this manuscript substantially expands the MIDL 2025 proceedings version \citep{gao2025plaxef}. The main extensions include:
\begin{itemize}
    \item \textbf{Multi-view EF estimation:} a new evaluation subset requiring both PLAX and A4C within each study, and a simple late-fusion strategy with improved accuracy (Section~\ref{subsec:multiview}).

    \item \textbf{Expanded artifact release and usage documentation:} we further release Hugging Face model checkpoints, a Hugging Face Space demo, and a Google Colab notebook enabling convenient inference and evaluation with the released models and label files (Section~\ref{sec:artifacts}). We also provide a concise artifact usage guide (Appendix~\ref{app:artifacts_usage}) to support reproducibility and ease of adoption.

    \item \textbf{Expanded methodological details for reproducibility:} we add implementation and hyperparameter details for the image-based bootstrap view classifier, the video view classifier, the A4C EF regressor, and the PLAX EF regressors in Appendix~\ref{app:training_details}.

    \item \textbf{Expanded clinical/technical context and literature:} a new Background section (Section~\ref{sec:background}) and a new Related Work section (Section~\ref{sec:related_work}) to better contextualize EF measurement, PLAX/A4C view characteristics, multi-view learning, and proxy supervision.
\end{itemize}

\paragraph{Organization.}
Section~\ref{sec:background} summarizes EF measurement and the clinical role of A4C/PLAX views. Section~\ref{sec:related_work} reviews prior work on EF estimation, PLAX modeling, multi-view echocardiography, and learning with scarce/noisy labels. Section~\ref{sec:method} details our dataset generation pipeline and model training. Section~\ref{sec:results} reports experimental results, and Section~\ref{sec:artifacts} summarizes released artifacts for reproducibility.

% =========================
% 2) BACKGROUND
% =========================
\section{Background}
\label{sec:background}

This section provides clinical and technical background on EF measurement and the echocardiographic views most relevant to this work. We separate this material from Related Work to reduce redundancy and to clarify which aspects are established clinical practice versus new machine-learning contributions.

\subsection{Left ventricular ejection fraction and its measurement}
Left ventricular ejection fraction (EF) is defined as the fraction of blood ejected from the left ventricle (LV) during systole:
\begin{equation}
    \mathrm{EF} \;=\; \frac{V_{\mathrm{ED}} - V_{\mathrm{ES}}}{V_{\mathrm{ED}}}\times 100\%,
\end{equation}
where $V_{\mathrm{ED}}$ and $V_{\mathrm{ES}}$ are end-diastolic and end-systolic LV volumes, respectively. In guideline-recommended transthoracic echocardiography, LV volumes are commonly estimated via the (bi)plane method-of-disks (Simpson's rule), using endocardial contours traced in apical long-axis views (typically A4C and A2C) \citep{lang2015ase}.

In contrast, PLAX-based EF estimation in clinical practice is frequently indirect and relies on geometric assumptions, for example by using linear LV internal dimensions (e.g., LVID) with analytic volume approximations such as the Teichholz formula \citep{teichholz1976}. While such formulas can be fast and interpretable, they can be sensitive to image quality, view alignment, and regional wall motion abnormalities that violate modeling assumptions. These limitations motivate end-to-end video-based learning approaches that attempt to map cine dynamics directly to EF without explicit intermediate measurements.

\subsection{A4C and PLAX views}
\begin{figure*}[t]
    \centering
    \begin{subfigure}{0.48\textwidth}
        \centering
        % Download and include a CC-licensed schematic locally (recommended).
        \includegraphics[width=\linewidth]{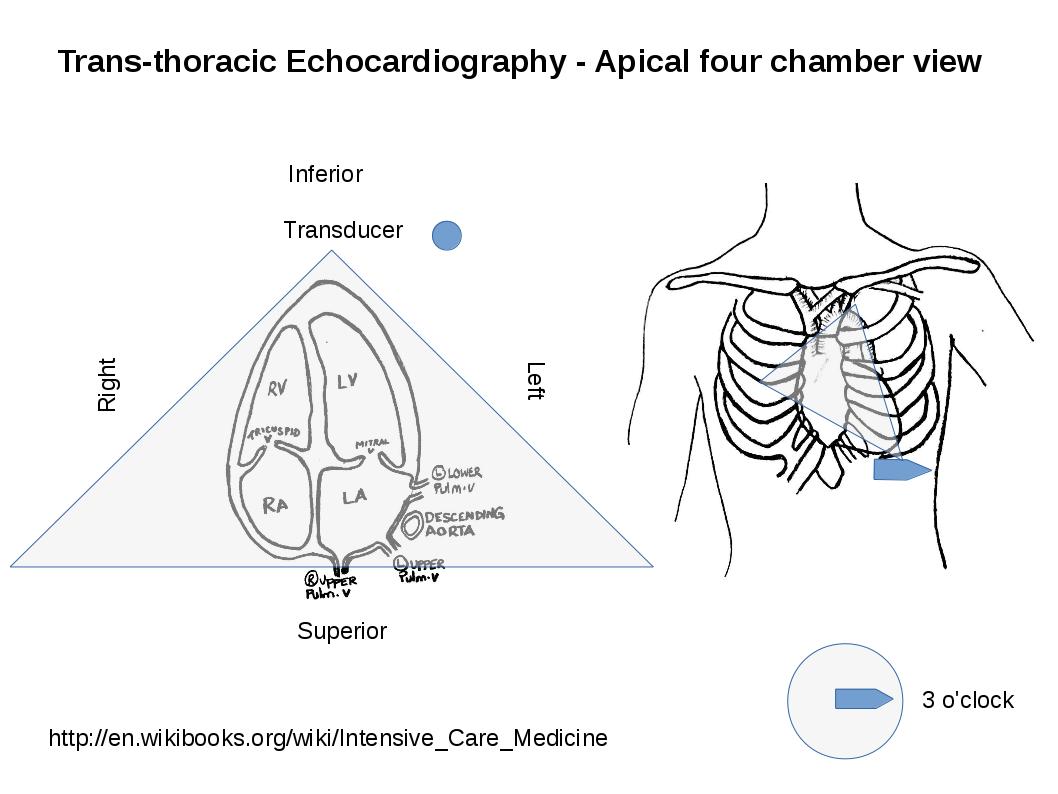}
        \caption{Apical four-chamber (A4C) schematic.}
        \label{fig:bg_a4c}
    \end{subfigure}
    \hfill
    \begin{subfigure}{0.48\textwidth}
        \centering
        \includegraphics[width=\linewidth]{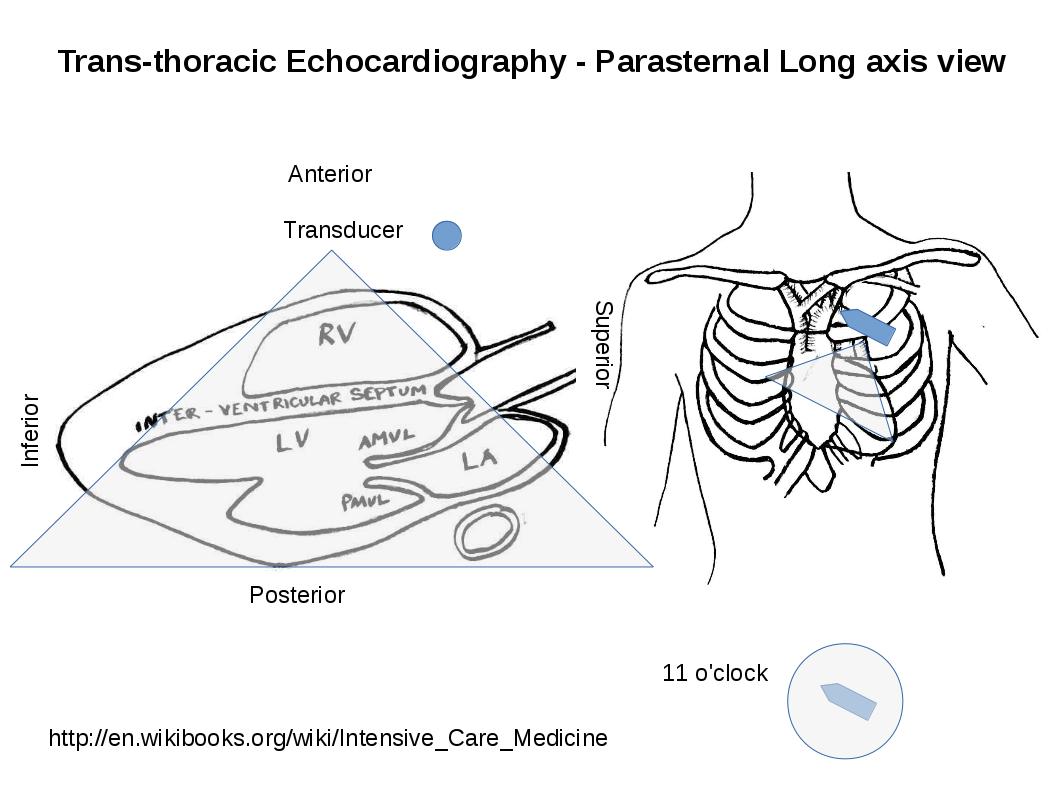}
        \caption{Parasternal long-axis (PLAX) schematic.}
        \label{fig:bg_plax}
    \end{subfigure}
    \caption{Standard long-axis echocardiographic views most relevant to this work. The schematics are reproduced from Wikimedia Commons under CC BY-SA 4.0 \citep{balaji_a4c_cc,balaji_plax_cc}.}
    \label{fig:bg_views}
\end{figure*}

A standard echocardiography study contains multiple views that capture complementary anatomy. This work focuses on the apical four-chamber (A4C) and parasternal long-axis (PLAX) views:
\begin{itemize}
    \item \textbf{A4C} provides a long-axis plane through both ventricles and atria, enabling volumetric assessment of the LV cavity and supporting guideline-recommended volume-based EF measurement \citep{lang2015ase}. A schematic illustration of the A4C view is shown in Fig.~\ref{fig:bg_a4c}.
    \item \textbf{PLAX} provides a parasternal long-axis slice through the LV and left atrium, with prominent visualization of the LV outflow tract and mitral valve apparatus. PLAX is often robust for rapid qualitative LV function assessment and is commonly available in point-of-care workflows \citep{russell2022plaxIsolation}. A schematic illustration of the PLAX view is shown in Fig.~\ref{fig:bg_plax}.
\end{itemize}

\subsection{Public data resources and supervision gaps}
A key challenge for PLAX EF modeling is that \emph{public} supervision is misaligned with acquisition frequency: PLAX is common in practice, yet public PLAX datasets rarely provide EF labels. Table~\ref{tab:bg_datasets} summarizes representative public resources and the type of supervision they provide.

\begin{table*}[t]
\centering
\caption{Representative public echocardiography resources used or referenced in this work and their supervision.}
\label{tab:bg_datasets}
\begin{tabular}{@{}lcc@{}}
\toprule
\textbf{Resource} & \textbf{View(s)} & \textbf{Supervision} \\
\midrule
EchoNet-Dynamic \citep{echonet-dynamic} 
& A4C 
& EF, volumes \\

EchoNet-LVH \citep{echonet-LVH} 
& PLAX 
& LVH-related labels \\

CAMUS \citep{leclerc2019camus} 
& A4C, A2C 
& Segmentation, volumes, EF \\

MIMIC-IV-Echo \citep{mimic_iv_echo,physionet} 
& multi-view 
& none (raw video) \\

MIMIC-IV-Note \citep{mimic_iv_note} 
& text 
& discharge notes \\

TMED-2 \citep{huangTMED2Dataset2022} 
& multi-view (images) 
& view labels \\
\bottomrule
\end{tabular}
\end{table*}

\subsection{Study-level versus video-level targets}
Clinical EF is typically reported at the \emph{study} level, whereas ML datasets often provide labels per video clip. Real-world studies contain multiple clips per view and varying image quality. Averaging predictions across clips within a study can reduce variance and can better match the clinical reporting unit. This motivates our study-level evaluation protocol and our multi-view fusion strategy, both performed at the study level (Section~\ref{sec:results}).

% =========================
% 3) RELATED WORK
% =========================
\section{Related Work}
\label{sec:related_work}

We review related work in four areas: (i) EF estimation from apical echocardiography (with emphasis on A4C), (ii) PLAX and non-apical EF estimation, (iii) multi-view learning in echocardiography, and (iv) learning with scarce and noisy supervision via proxy labels and report mining.

\subsection{EF estimation from apical echocardiography}
Most machine-learning work on echocardiographic EF estimation targets apical views because they support standardized volumetric measurements and are the dominant source of labeled EF in clinical reporting. EchoNet-Dynamic popularized end-to-end EF regression from A4C cine videos and demonstrated that spatiotemporal deep networks can achieve clinically competitive error when trained on curated EF labels \citep{echonet-dynamic}. Typical pipelines map a fixed-length clip (or multiple sampled clips) to a scalar EF prediction and may incorporate temporal aggregation to stabilize predictions. Beyond direct regression, several approaches estimate EF through intermediate structure segmentation and volume computation, leveraging the method-of-disks or related geometric constructs; large-scale datasets such as CAMUS provide ED/ES annotations for segmentation and derived indices including EF \citep{leclerc2019camus}.

Another practical concern is that apical views can be compromised by foreshortening, drop-out, and limited endocardial definition. Real-time systems have been proposed to jointly address EF estimation and view-quality failure modes (e.g., foreshortening detection) to improve reliability in deployment settings \citep{smistad2020realtime}. Collectively, these works show that A4C-based EF prediction is feasible and can be highly accurate \emph{when} curated labels and consistent view definitions are available---a condition that does not hold for PLAX EF in public data.

\subsection{PLAX and non-apical EF estimation}
PLAX is routinely acquired and frequently used in focused/point-of-care examinations, and clinical studies suggest that PLAX alone can be informative for identifying LV systolic dysfunction \citep{russell2022plaxIsolation}. However, PLAX is not commonly the primary view for EF reporting, resulting in a scarcity of PLAX--EF supervision.

Existing PLAX EF efforts often follow one of two paradigms. 

\textbf{(i) Indirect measurement pipelines:} EF is estimated from PLAX via intermediate surrogates such as linear LV dimensions (fractional shortening, LVID-based volume approximations) or landmark-based measurement extraction \citep{teichholz1976}. Recent work proposed lightweight landmark detection on PLAX to estimate EF from sparse annotations, effectively learning to recover measurement-derived EF values \citep{10.1117/12.2611239}. Such methods benefit from interpretability but introduce additional error sources, including landmark uncertainty and inter-observer measurement variability.

\textbf{(ii) Direct PLAX EF prediction:} Proprietary or closed systems have reported PLAX EF performance without sufficient disclosure for independent reproduction. For example, ExoAI reported PLAX-specific error in a retrospective evaluation, but training data and algorithmic details were not fully available to the community \citep{diagnostics14161719}. More recently, point-of-care studies have explored PLAX-based EF estimation via motion/wall tracking and related approaches \citep{vega2025walltracking}. Overall, the field lacks a widely reproducible, public PLAX--EF benchmark with runnable artifacts, limiting systematic comparisons across methods.

\subsection{Multi-view learning in echocardiography}
Echocardiography is inherently multi-view: a single study typically includes multiple standardized planes (parasternal and apical views) that emphasize different anatomy and motion patterns. Multi-view learning has therefore been explored across tasks including view classification, segmentation, disease classification, and quantitative function estimation. Fusion strategies range from \emph{late fusion} (combining independently trained view-specific predictors) to learned fusion via attention or global-local aggregation modules \citep{zheng2023glfusion}.

For EF-related tasks, multi-view modeling is appealing because different views fail for different reasons (e.g., apical foreshortening versus parasternal drop-out). Prior clinical ML systems have adapted EF estimation to work from a subset of commonly acquired point-of-care views (including PLAX and A4C), and to leverage whichever views are available \citep{asch2019autoef,asch2021autoef_pocus}. Recent architectures explicitly ingest multiple standard views via multi-stream encoders and transformer-based fusion, often combining appearance and motion cues (e.g., optical flow) to improve robustness \citep{alven2024multistream}. Additionally, transformer-based multi-view encoders have been developed to aggregate view-level representations into study-level embeddings for downstream tasks, highlighting the broader trend toward study-level, multi-view representations \citep{tohyama2025multiviewembedding}. Our work contributes to this line by providing a reproducible A4C--PLAX multi-view EF test subset and demonstrating that even simple unweighted late fusion yields measurable gains when both views are present.

\subsection{Learning with scarce and noisy labels: proxy supervision and report mining}
Label scarcity and label noise are pervasive in clinical ML. When the desired supervision is not routinely recorded for a given view or modality, a common approach is to use \emph{proxy labels} or \emph{pseudo-labels} from a teacher model (self-training), enabling large-scale training despite limited ground truth \citep{lee2013pseudo,xie2020noisy}. Weak supervision frameworks further formalize the combination of multiple noisy sources (heuristics, rules, imperfect models) to create training labels at scale \citep{ratner2016snorkel}.

In echocardiography, EF is often present in unstructured reports rather than machine-readable fields. Prior work has extracted EF values from clinical text using NLP pipelines, enabling retrospective cohort construction from notes \citep{kim2017efextraction}. More recently, large language models have been used as flexible extractors of structured information from heterogeneous clinical narratives; in our pipeline we use GPT-4-based extraction with subsequent validation against video-model predictions to improve reliability \citep{openai2024gpt4technicalreport}. Combining proxy-supervised training with an \emph{independent} note-derived ground-truth evaluation cohort allows us to (i) learn PLAX EF models at scale and (ii) transparently quantify performance without circularly evaluating on the teacher-generated labels.

\section{Dataset Generation and Model Training}
\label{sec:method}
\textit{The key step of our methodology—and the major challenge—was creating a~labeled dataset for PLAX from available data, despite the scarcity of publicly available PLAX-specific labels. Because PLAX images and EF labels are not routinely paired in existing repositories, we developed novel techniques to both identify PLAX views and assign approximate EF values, effectively circumventing the lack of direct ground-truth annotations.}\\

\begin{figure}[!t]
\centerline{\includegraphics[width=\columnwidth]{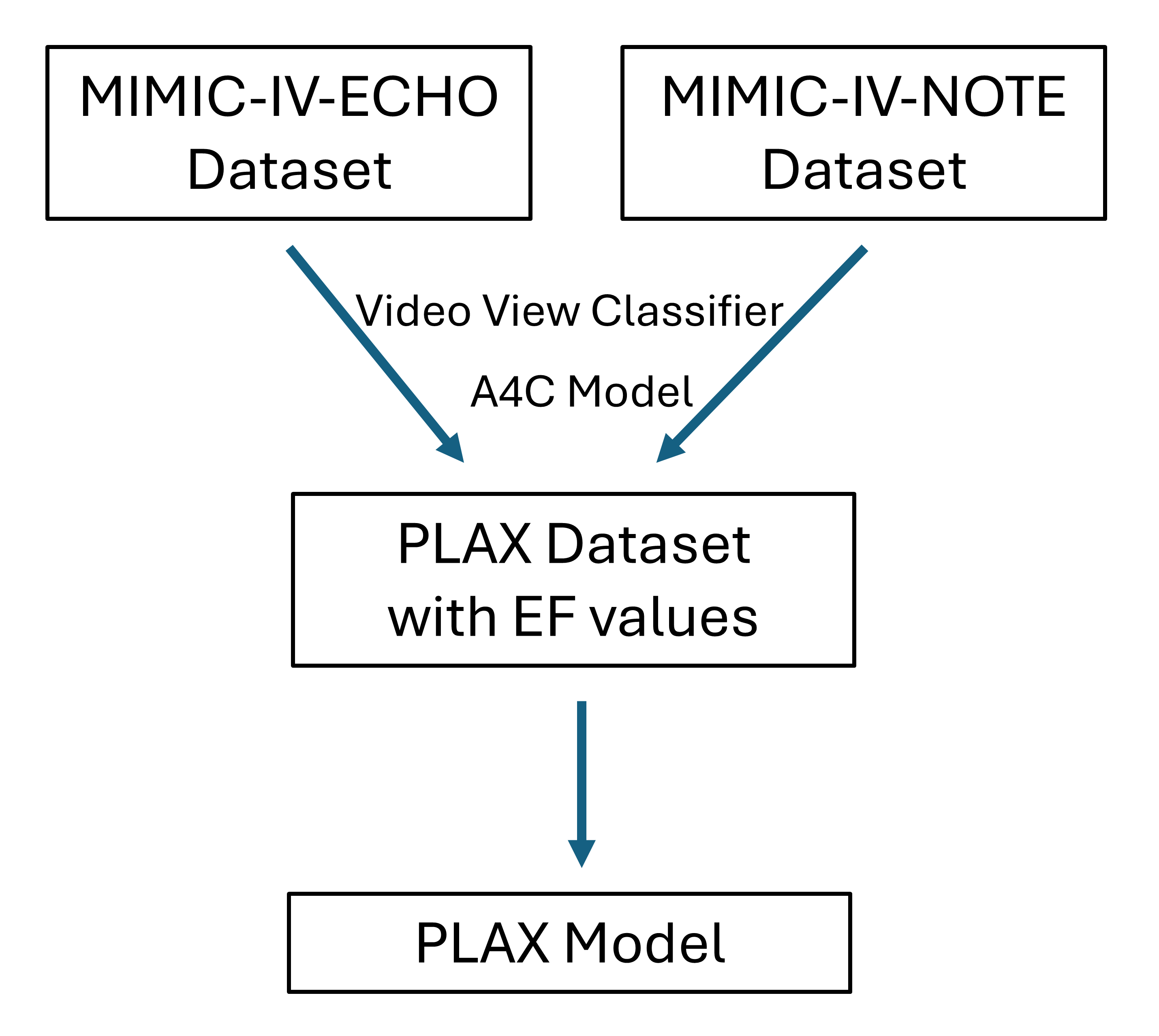}}
\caption{High-Level PLAX EF Prediction Pipeline.}
\label{fig1}
\end{figure}

Two major datasets from PhysioNet\citep{physionet} were utilized for this study:
\begin{itemize}
    \item \textbf{MIMIC-IV-Echo}\citep{mimic_iv_echo}: Contains approximately 500k echocardiographic videos without view type labels.
    \item \textbf{MIMIC-IV-Note}\citep{mimic_iv_note}: Includes around 331k discharge notes, of which only a subset contains EF values in unstructured text.

\end{itemize}

The goal was to extract PLAX view videos with EF data and use them to train a machine learning model for EF prediction. This task faced two main challenges: first, the MIMIC-IV-Echo dataset lacked labels for echocardiographic view types, requiring the development of a video view classifier to identify PLAX views; second, the MIMIC-IV-Echo and MIMIC-IV-Note datasets were not directly linked, making it difficult to associate discharge notes with corresponding echocardiographic studies. Even after applying time-based correlations, the number of valid note-study pairs remained insufficient for training a robust PLAX model.

Consequently, we needed to leverage most studies in the MIMIC-IV-Echo dataset to generate training data for PLAX EF prediction. Specifically, we first trained a video view classifier to identify A4C and PLAX views within the dataset. Next, we trained an A4C model using a publicly available dataset to estimate EF values. These EF predictions were then applied as proxy labels for PLAX videos within the same study, enabling the development of a PLAX-specific model. A high-level overview is illustrated in Fig. \ref{fig1}.

\subsection{Video View Classifier Training}  \label{VideoViewClassifier}
A~classifier capable of distinguishing echocardiographic views into A4C, PLAX, and "OTHER," was essential for accurately selecting A4C and PLAX videos from the MIMIC-IV-Echo dataset. 

We first leveraged the \textbf{TMED-2} dataset \citep{huangTMED2Dataset2022}, which contains approximately 25,000 labeled echocardiographic images spanning a~range of views, including A4C, PLAX, apical two-chamber (A2C), parasternal short-axis (PSAX), and a~combined category of miscellaneous views labeled as “A4C/A2C/OTHER.” Using these labeled images, we trained a~ResNet-34 image classifier. 

The image classifier was applied frame-by-frame to MIMIC-IV-Echo videos, and aggregated predictions were used to assign video-level classifications. The model produced log-softmax outputs, which were converted to probabilities via the exponential transform, and the class with the highest score was taken as the predicted view.  

However, the MIMIC-IV-Echo dataset does not provide ground-truth labels for the videos, preventing direct measurement of the image classifier’s accuracy on this dataset. 

To evaluate performance on video data in the absence of ground-truth annotations, two independent reviewers manually assessed 300 randomly selected test videos. The evaluation results are summarized below:
\begin{itemize}
    \item For 100 videos predicted as A4C, two reviewers identified 91/87 as correct.
    \item For 100 videos predicted as PLAX, two reviewers identified 78/72 as correct.
    \item For 100 videos not predicted as A4C/PLAX, two reviewers identified 94/98 as correct.
\end{itemize}

\begin{figure*}[htbp]
  \centering
  \includegraphics[width=0.8\textwidth]{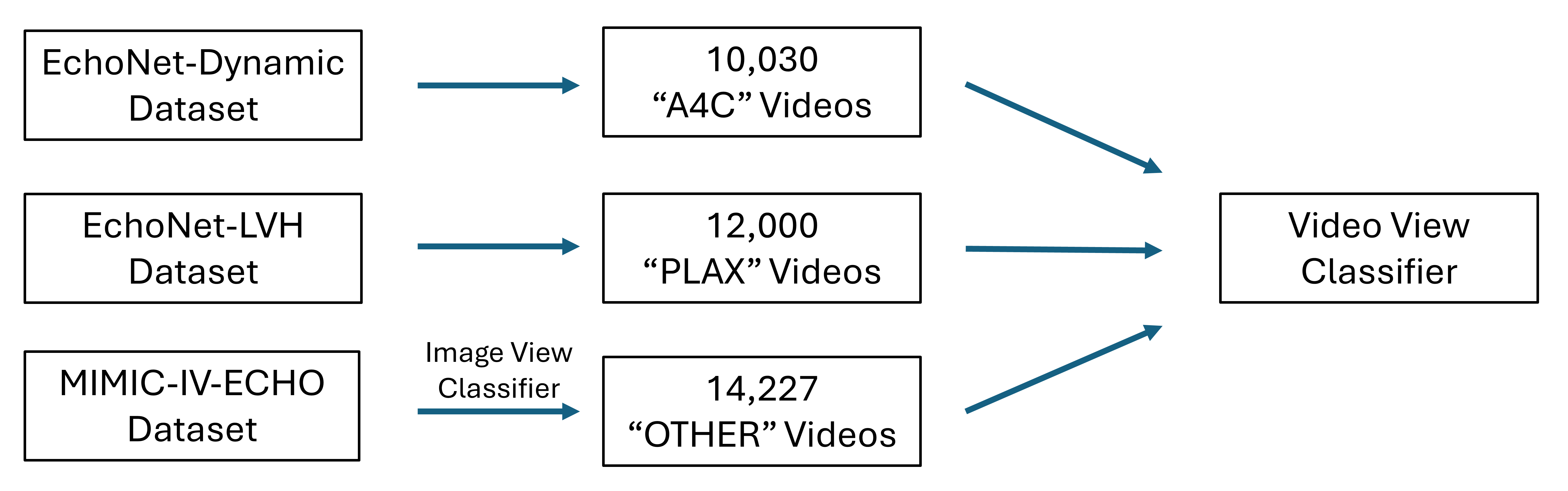}
  \caption{Multi-Dataset Strategy for Video View Classification.}
  \label{fig2}
\end{figure*}

These findings indicated that while the image classifier was highly reliable in identifying “OTHER” views, its accuracy for A4C and PLAX was insufficient for downstream use, motivating the development of a~more robust video-based classifier. To train such a~classifier, we required videos labeled as A4C, PLAX, and "OTHER" views. This process is outlined in the flowchart in Fig. \ref{fig2}.

For A4C and PLAX training data, we utilized the following two EchoNet datasets. EchoNet is a publicly available echocardiographic dataset designed for ML research. Unlike the MIMIC datasets, EchoNet datasets are highly curated collections of echocardiographic videos with specific annotations for view types and clinical parameters. 

\begin{itemize}
    \item \textbf{EchoNet-Dynamic}\citep{echonet-dynamic}: Approximately 10,000 videos labeled as A4C with EF values.
    \item \textbf{EchoNet-LVH}\citep{echonet-LVH}: Approximately 12,000 videos labeled as PLAX without EF values.
\end{itemize}

Constructing an “OTHER” video dataset was more challenging, as it required broad representation of alternative views. To address this, we leveraged the image classifier (which showed reliable performance for “OTHER”) to extract videos from MIMIC-IV-Echo. Specifically, we selected:  
\begin{itemize}
    \item 4,212 A2C videos (exp-transformed score $>0.6$),
    \item 4,015 PSAX videos (exp-transformed score $>0.6$),
    \item 6,000 videos labeled as "A4C/A2C/OTHER" (exp-transformed score $>0.9$).
\end{itemize}
The exp-transformed scores were manually set to balance the number of videos across different categories, ensuring adequate representation for training and minimizing contamination:

\begin{itemize}
    \item First, for better model performance, it’s desirable to balance A4C, PLAX, and “OTHER” categories, targeting “OTHER” dataset sizes similar to EchoNet-Dynamic (10,030 A4C) and EchoNet-LVH (12,000 PLAX). 

    \item Second, due to the limitations of the TMED dataset, the “OTHER” category was constructed using A2C, PSAX, and “A4C/A2C/OTHER” labels. Within “A4C/A2C/OTHER”, manual verification showed that increasing the exp-transformed score threshold reduced A4C/A2C contamination. Thus, we set 0.9 to ensure diversity while minimizing A4C inclusion. 
    \item Third, to balance A2C and PSAX within “OTHER”, while maintaining overall dataset proportionality between A4C, PLAX and “OTHER”, we set a 0.6 threshold for both A2C and PSAX. This approach allowed us to create a comprehensive "OTHER" dataset while maintaining diversity in the video views. 
\end{itemize}

Finally, the overall distribution of training data for the video view classifier was as follows:
\begin{itemize} 
\item \textbf{A4C:} 10,030 videos, sourced entirely from the EchoNet-Dynamic dataset. 
\item \textbf{PLAX:} 12,000 videos, sourced entirely from the EchoNet-LVH dataset. 
\item \textbf{"OTHER":} 14,227 videos, ML-classified from the MIMIC-IV-Echo dataset. 
\end{itemize}

For the final video view classifier, we utilized a~pretrained X3D-s model \citep{feichtenhofer2020x3dexpandingarchitecturesefficient} on this dataset. Compared with the ResNet-34 image classifier, the video-based model demonstrated substantially higher accuracy in distinguishing A4C and PLAX views, ensuring higher-quality data for subsequent analyses. This video-based classifier formed the backbone of our video view classification pipeline, achieving robust performance and enabling accurate selection of PLAX and A4C videos for downstream tasks. 

Additional implementation details for the view-classification models (image-based
bootstrapping on TMED-2, video-model fine-tuning, clip sampling, data
augmentation, optimization, and compute setup) are provided in
Appendix~\ref{app:view_classifier_training}.

\subsection{A4C Model Training}\label{2.2}
We train an A4C EF regression model on \textbf{EchoNet-Dynamic}, which contains 10,030 A4C cine videos with expert-reported EF labels. Following \citet{echonet-dynamic}, we implement a spatiotemporal \textbf{R(2+1)D} network (TorchVision \texttt{r2plus1d\_18}) initialized from a public video-pretrained checkpoint, replacing the final classification layer with a single-neuron regression head. Training and model selection follow the official EchoNet-Dynamic \texttt{TRAIN}/\texttt{VAL}/\texttt{TEST} split, with mean squared error (MSE) as the regression loss. 

The resulting model achieves an MAE of \textbf{4.37\%} on the test split, closely matching the 4.1\% MAE reported in \citet{echonet-dynamic}. Full implementation details (clip sampling, preprocessing/augmentation, optimization, and checkpoint selection) are provided in Appendix~\ref{app:train_a4c}. This trained A4C predictor is subsequently used as a teacher model to generate study-level proxy EF labels for PLAX training (Section~\ref{2.5}).

\subsection{Ground Truth Data Generation}\label{2.3}

\begin{figure}[htbp]
\centerline{\includegraphics[width=\columnwidth]{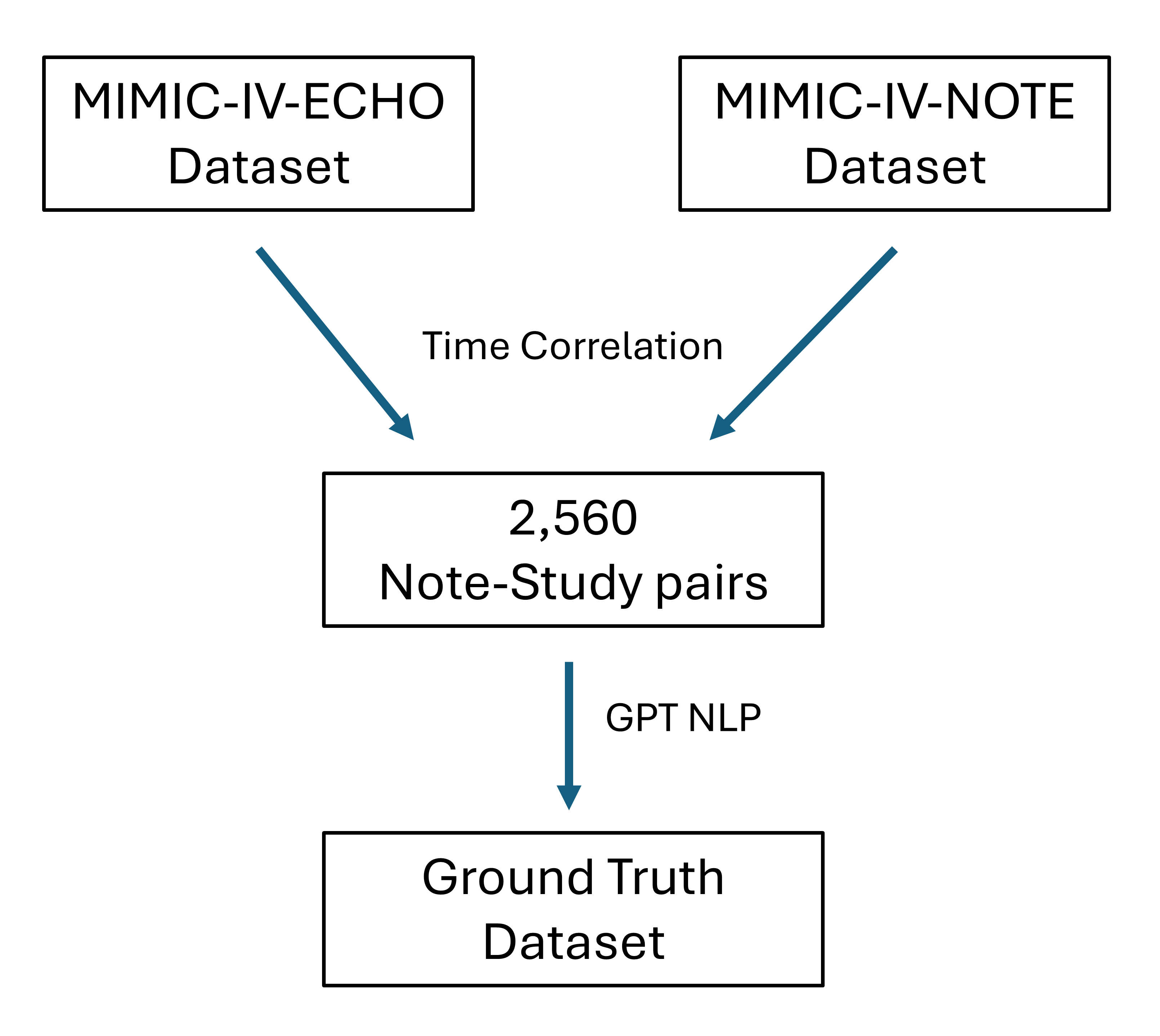}}
\caption{Ground Truth Dataset Generation Flow Chart.}
\label{fig3}
\end{figure}

A~ground truth dataset was essential for evaluating the true error of the PLAX EF model. Because the PLAX EF model was trained on EF values indirectly generated by the A4C model, its performance inherently included the compounded error of A4C predictions. To measure the error of the PLAX EF model independently, we constructed a~ground truth dataset with directly validated EF values. An overview of this process is shown in Fig.~\ref{fig3}.

No direct mapping exists between studies in the MIMIC-IV-Echo and MIMIC-IV-Note datasets. To bridge this gap, we performed a~time-based correlation between the two datasets using the same patient ID. This approach identified 2,560 note-study pairs where echocardiography studies and clinical notes were recorded within a~60-day window.

Given the unstructured and messy text format of the notes, we utilized the GPT-4 API\citep{openai2024gpt4technicalreport} to extract EF values and other relevant information. The API was instrumental in parsing and cleaning the free-text notes to retrieve meaningful clinical data, which was also publicly available (Section~\ref{sec:artifacts}). This automated parsing and cleaning process yielded 921 valid note-study pairs within the 60-day window with a valid EF value.

To assess whether these note-derived EF values are reliable rather than artifacts of a single extractor, we repeated the extraction with two additional, independent and more recent large language models (GPT-5.5 and Claude Opus 4.8) using an identical prompt. Across the notes for which all three models returned a numeric EF, the three models agreed exactly in 99.5\% of cases, with pairwise Pearson correlations $r \approx 1.0$. These results support the reproducibility of the note-derived EF extraction and reduce the likelihood that the evaluation labels are driven by model-specific extraction artifacts. For consistency with the originally released benchmark and to avoid changing the evaluation protocol, we retained the GPT-4-extracted EF values as the final labels.

Next, we applied the video view classifier from Section~\ref{VideoViewClassifier}
to identify studies containing at least one candidate A4C clip and one candidate
PLAX clip. After this view-based filtering, 902 note--study pairs remained.
We then enforced a confidence threshold by requiring an exp-transformed score
greater than 0.5 for both the A4C and PLAX predictions, yielding 848
high-confidence pairs.

To validate the note-extracted EF values, we ran the A4C model described in Section~\ref{2.2} on the A4C videos for the studies and compared the predictions with the note labels, obtaining a~study-level MAE of 7.82\%. To further improve accuracy, we restricted the correlation to a~1-day window between notes and studies. This yielded 295 high-confidence note–study pairs, each containing at least one PLAX and one A4C video. In addition, videos were screened to ensure a~duration greater than one second and to exclude studies containing color Doppler imaging. After this filter, the final test set included 290 studies with 1,017 A4C videos and 295 studies with 1,320 PLAX videos (note that the study counts overlap).

On this high-confidence cohort, the A4C model achieved a~MAE of 6.95\% when compared against note-extracted labels. This closely matches the reported out-of-sample performance of 6.0\% in \citep{echonet-dynamic}, providing evidence that the note-derived EF values are reliable. These 295 studies (comprising 1,017 A4C videos and 1,320 PLAX videos) formed our ground truth test set, and their labels are publicly available (Section~\ref{sec:artifacts}).  

Although this ground truth dataset was not large enough to train the PLAX EF model, it serves as an~independent benchmark. The error measured on this set provides a~robust evaluation of the PLAX EF model that does not depend on proxy EF values generated by the A4C model.

\subsection{View Classifier Fine-Tuning}

The video view classifier is essential to our pipepline. To further enhance the performance of our video view classifier (X3D model), we leveraged the 902 valid note-study pairs identified within the 60-day window as described in Section~\ref{2.3}. These pairs were used to further refine the classifier's ability to identify A4C views.

\begin{figure}[!t]
\centerline{\includegraphics[width=\columnwidth]{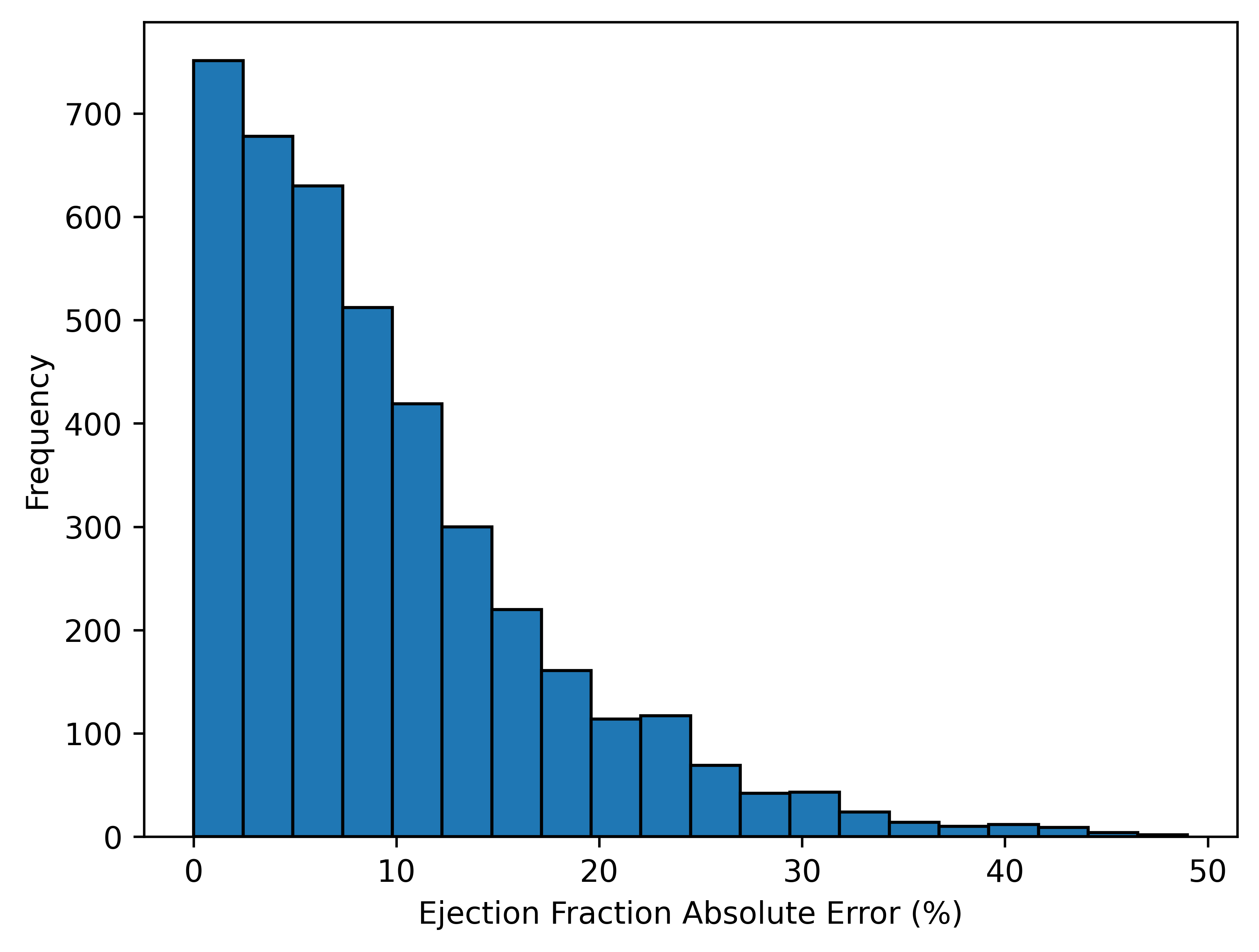}}
\caption{Error distribution of the A4C model within 60-day window.}
\label{fig4}
\end{figure}

First, we applied the view classifier to extract A4C videos from studies, identifying a~total of 4,131 videos within the 60-day window. Next, we ran our A4C model on these identified A4C videos, producing the error distribution shown in Fig. \ref{fig4}.

Upon reviewing the 10 videos with the highest errors, we found that their views were either not true A4C views or partial A4C views. This observation demonstrated that the errors of the A4C model could be effectively utilized to fine-tune the video view classifier.

To address this, we fine-tuned the video view classifier using a~subset of videos with extreme errors:
\begin{itemize}
    \item Videos with MAE $<3\%$ were labeled as A4C.
    \item Videos with MAE $>20\%$ were labeled as "OTHER."
\end{itemize}
This subset comprised 1,346 videos, which were split equally into training and test sets for fine-tuning.

The X3D model was fine-tuned using the labeled training set, with the objective of improving its ability to distinguish between A4C and "OTHER" views. After fine-tuning, the test set was regenerated using the fine-tuned classifier, and the MAE was recalculated using the A4C model. The MAE was reduced from 6.83\% to 5.14\%, demonstrating a~significant improvement in classifier accuracy.

This fine-tuning process enabled the classifier to more accurately identify A4C views, reducing misclassifications and ensuring higher-quality input for downstream tasks.

\subsection{PLAX Dataset Generation}\label{2.5}

\begin{figure}[htbp]
\centerline{\includegraphics[width=0.5\columnwidth]{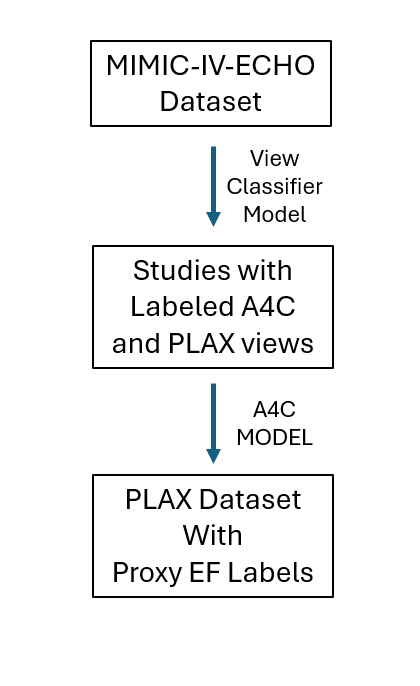}}
\caption{Overview of the PLAX Dataset Construction Pipeline Using Proxy Supervision.}
\label{fig_plax_pipeline}
\end{figure}

We applied the video view classifiers, both before and after fine-tuning, to the MIMIC-IV-Echo dataset, where the patients appearing in the ground truth dataset were excluded. An overview of this process is shown in Fig.~\ref{fig_plax_pipeline}.

\begin{figure*}[t]
    \centering
    \begin{subfigure}{0.4\linewidth}
        \centering
        \includegraphics[width=\linewidth]{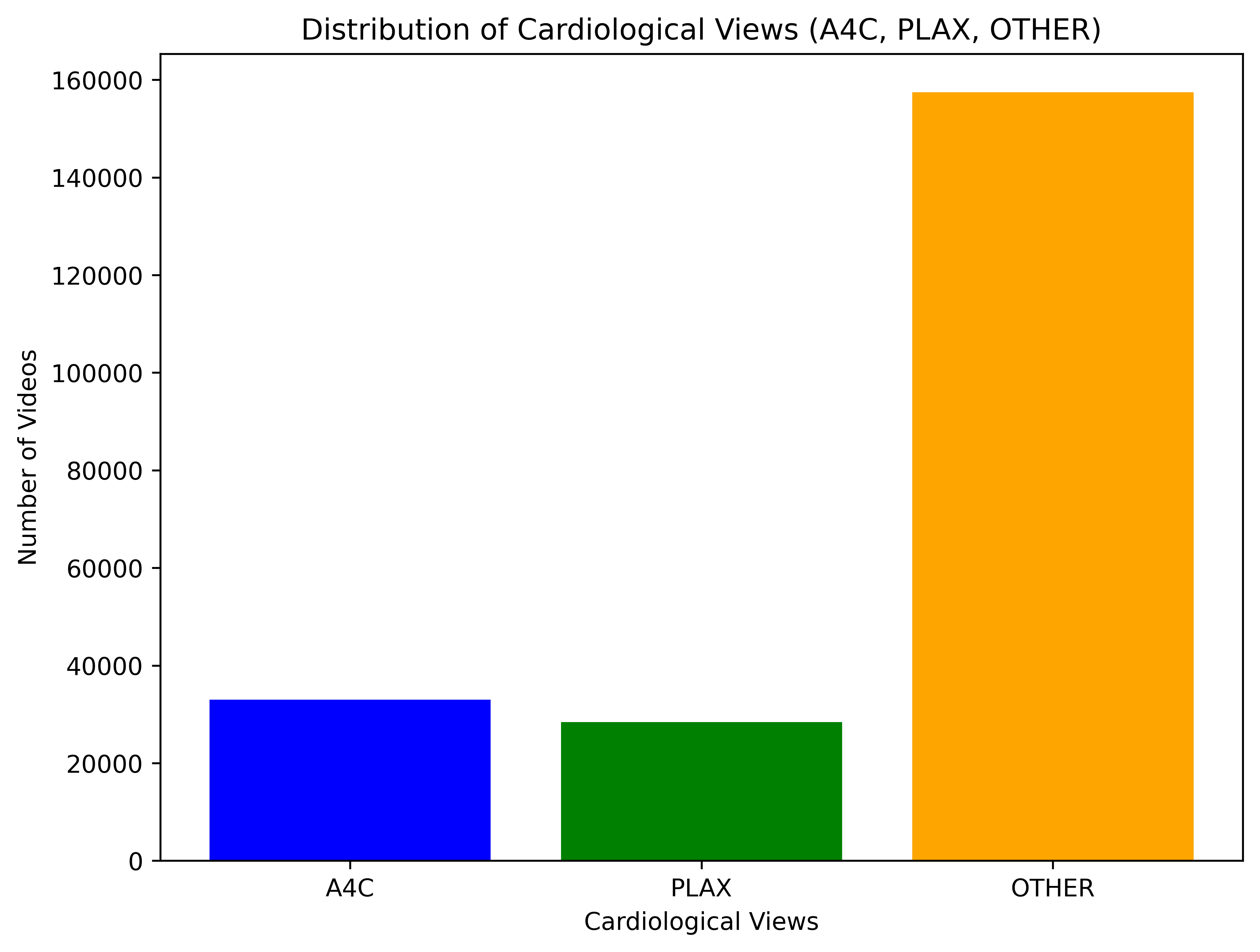}
        \caption{Label distribution before fine-tuning.}
        \label{fig:pre_finetune}
    \end{subfigure}
    \hfill
    \begin{subfigure}{0.4\linewidth}
        \centering
        \includegraphics[width=\linewidth]{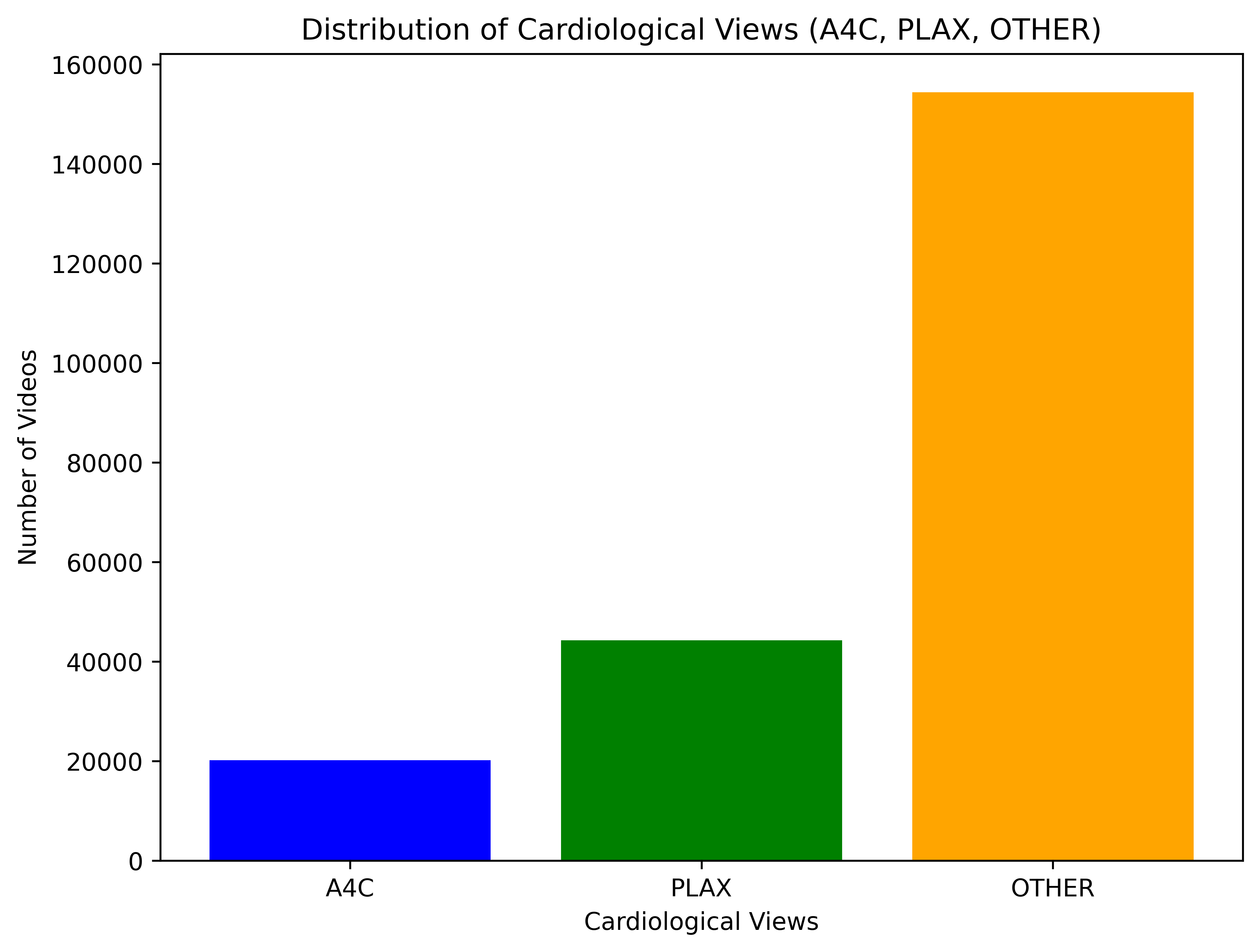}
        \caption{Label distribution after fine-tuning.}
        \label{fig:post_finetune}
    \end{subfigure}
    \caption{Comparison of label distributions before and after fine-tuning the video view classifier.}
\end{figure*}

The label distribution of videos before fine-tuning is shown in Fig. \ref{fig:pre_finetune}, while the distribution after fine-tuning is shown in Fig. \ref{fig:post_finetune}. After fine-tuning, the classifier identified A4C views more strictly, reducing the number of videos labeled as A4C by approximately 15,000, resulting in a~total of around 20,000 videos. Conversely, the number of videos classified as PLAX views increased significantly.

To ensure maximum accuracy, we selected PLAX views from the pre-fine-tuning classifier and A4C views from the post-fine-tuning classifier, eliminating any overlapping videos. Most studies contained both an~A4C video and a~PLAX video. For A4C videos in each study, we used the A4C model described in Section~\ref{2.2} to generate EF values, and then averaged those values to serve as the EF label for the PLAX videos in the study.

The final PLAX training dataset consisted of 25,532 videos, comprising 4,822 studies (80\% training, 20\% validation). Different videos within the same study were assigned to the same split for a clean validation set. Labels are publicly available (Section~\ref{sec:artifacts}).

% \begin{figure}[H]
%     \centering
%     \includegraphics[width=0.4\linewidth]{D2.jpg}
%     \caption{PLAX Training Dataset Generation Flow Chart.}
% \end{figure}
For testing, we used the ground truth dataset described in Section~\ref{2.3}. Using the pre-fine-tuned view classifier, which is stricter in identifying PLAX views, we extracted 1,320 PLAX videos from the 295 studies. To ensure strict data separation, patients included in the test set were excluded from both the training and validation sets.

With the proxy-labeled training set and the held-out ground-truth cohort in
place, we next train PLAX EF models and report quantitative results.

\section{Experimental results and analyses}
\label{sec:results}

Building on the proxy-labeled PLAX training set constructed in
Section~\ref{2.5} and the independent note-derived ground-truth cohort described
in Section~\ref{2.3}, we now evaluate PLAX-based EF prediction and quantify the
benefits of multi-view integration. Unless otherwise stated, we report
\emph{study-level} performance by averaging video-level predictions across all
clips of a given view within each study, which better matches the clinical unit
of EF reporting (Section~\ref{sec:background}). We first present single-view PLAX
results, and then analyze a simple late-fusion strategy that combines A4C and
PLAX predictions on studies containing both views.

\subsection{PLAX Model Training and Results}\label{sec:plax-results}

We trained two R(2+1)D models with different configurations based on the PLAX training dataset mentioned in section~\ref{2.5}. The training details and configurations are summarized in Table~\ref{tab:model_training}.

\begin{table}[htbp]
  \centering
  \caption{Training configurations for PLAX EF prediction models.}
  \label{tab:model_training}
  \begin{tabular}{l|c|c|c}
    \bfseries Model & \bfseries Batch Size & \bfseries MAE &  \bfseries Correlation \\ 
    R(2+1)D & 16 & 6.93\% & 0.659\\ 
    R(2+1)D & 32 & 6.89\% & 0.656\\ 
  \end{tabular}
\end{table}

Both models were trained for 100 epochs using a~learning rate (LR) scheduler with an~initial LR of 0.001, a~patience of 5 epochs and a~reduction factor of 0.1. Input clips were spatially standardized using padding and random cropping to a fixed resolution. Standard preprocessing, including padding and random cropping, was applied. For each configuration, the checkpoint with the lowest validation error was selected as the final model. Additional training and implementation details for the PLAX models are provided in Appendix~\ref{app:plax_training_details}.

At test time, study-level predictions were obtained by averaging video-level predictions, and MAE was computed per study. The final EF estimate was produced by taking an~unweighted average of the two model outputs on the ground truth dataset described in Section~\ref{2.3}, yielding an~MAE of \textbf{6.86\%}.

\begin{figure}[htbp]
\centerline{\includegraphics[width=\columnwidth]{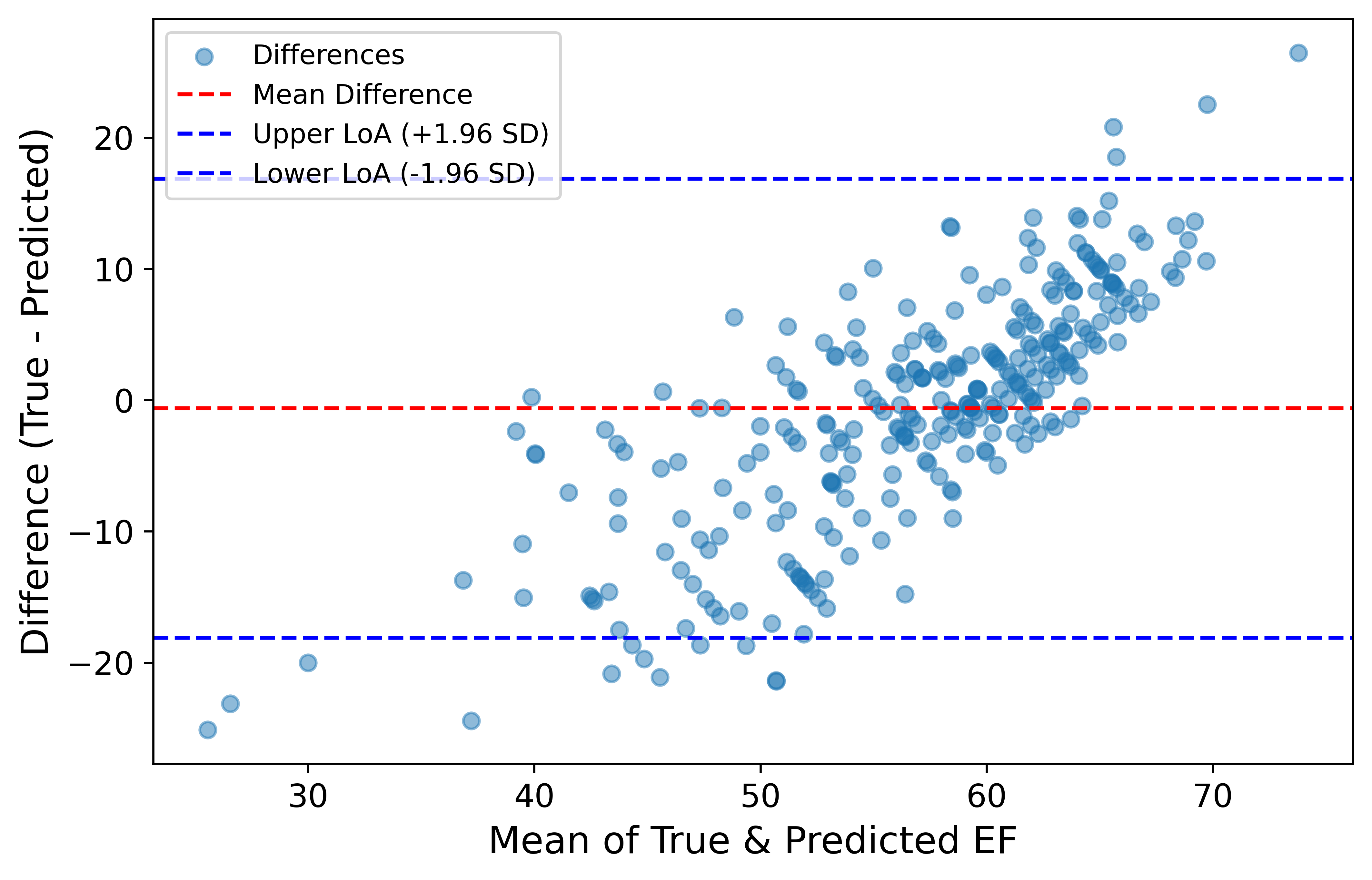}}
\caption{Scatter Plot for PLAX EF prediction. The dashed red line indicates the line of identity.}
\label{fig7}
\end{figure}

\begin{figure}[htbp]
\centerline{\includegraphics[width=\columnwidth]{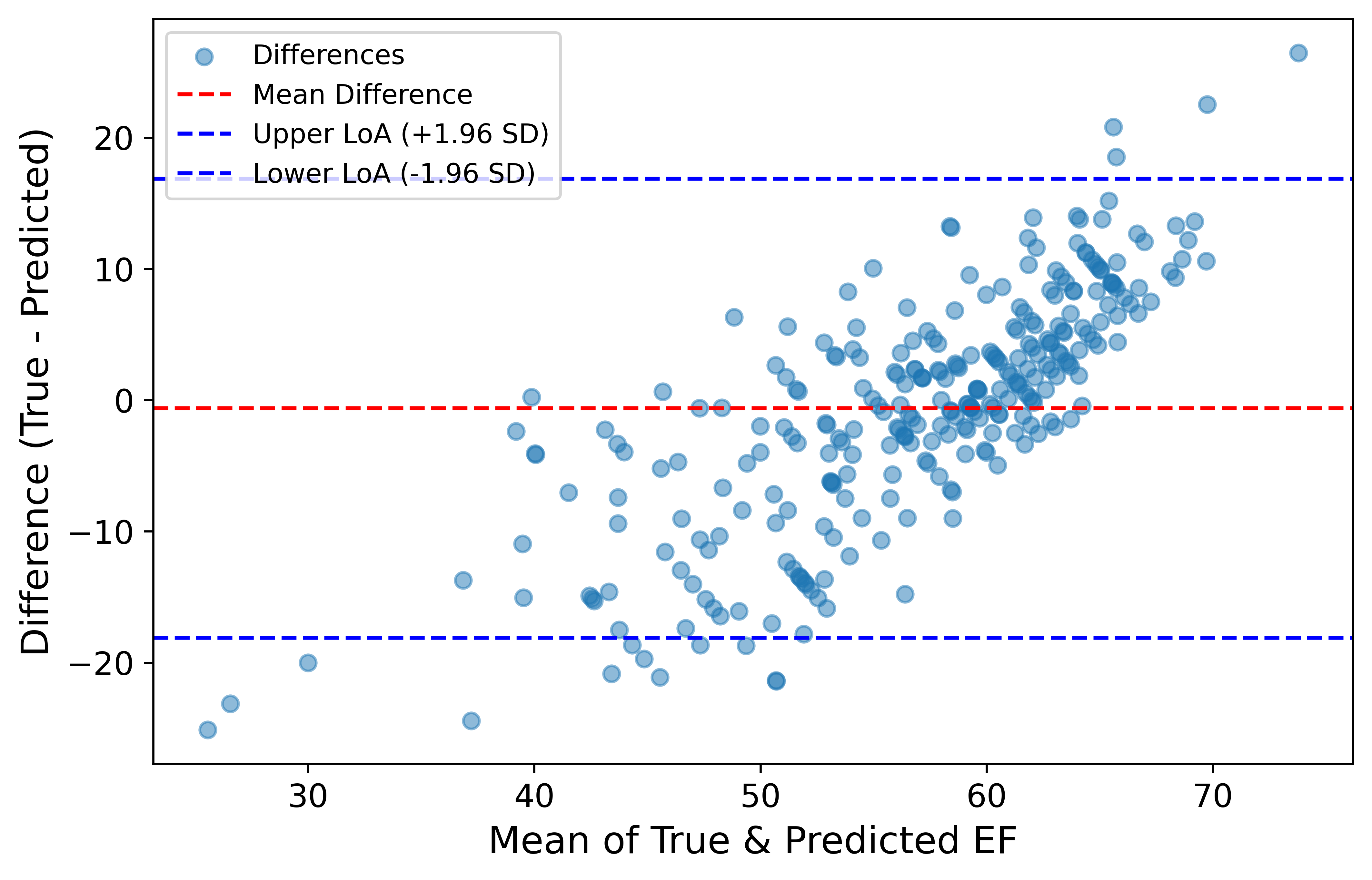}}
\caption{Bland-Altman Plot for PLAX EF prediction.}
\label{fig8}
\end{figure}
To further assess the agreement between predicted and true EF values, we computed the Pearson correlation coefficient of 0.670, indicating a reliably positive correlation between our model’s predictions and the ground truth. The scatter plot (Fig. \ref{fig7}) demonstrates that predictions generally follow the line of identity, though variance increases for higher EF values. Additionally, we present a Bland-Altman plot (Fig. \ref{fig8}) to analyze systematic bias and agreement limits. The mean difference (bias) is -0.62\%, suggesting no significant systematic offset in predictions. The upper and lower limits of agreement (LoA) are 16.87\% and -18.10\%, respectively, indicating some variability in the prediction errors. Notably, the plot shows increased dispersion at higher EF values, which aligns with previous observations in EF estimation models.

One potential explanation for this variance and the moderate correlation is the use of proxy labels derived from an A4C-based EF model, which can introduce compounded errors. Additionally, since the A4C model was not originally trained on MIMIC datasets, its predictions can introduce some domain shift. However, despite these challenges, our work provides, to the best of our knowledge, the first \emph{publicly reproducible} PLAX EF benchmark with disclosed methodology and runnable inference artifacts. On this cohort, our PLAX model achieves lower MAE than previously reported PLAX EF results we are aware of, including proprietary systems with limited disclosure and indirect measurement-based pipelines. Prior works either lack methodological transparency (e.g., ExoAI with an MAE of 7.29\%) \citep{diagnostics14161719} or rely on indirect EF estimation methods such as LVID measurement (MAE 8.45\%) \citep{10.1117/12.2611239}, which introduce inter-observer variability. By directly predicting EF from full PLAX cine videos, our approach avoids these manual dependencies while setting a reproducible standard for future research.

\subsection{Aggregating A4C and PLAX Views}\label{subsec:multiview}

Because most echocardiographic studies contain both A4C and PLAX views, we investigated whether integrating information across views could further improve EF estimation. The intuition is that the two perspectives capture complementary anatomical details, and combining them may reduce variance and correct for view-specific biases.  

We used the ground truth dataset described in Section~\ref{2.3}, which includes 290 studies with 1,017 A4C videos and 295 studies with 1,320 PLAX videos (with partial overlap between the study sets). To construct a~multi-view test set, we retained only studies containing at least one A4C video and one PLAX video. This yielded 284 studies comprising 1,000 A4C videos and 1,275 PLAX videos. These studies formed our multi-view test set, which is publicly available (Section~\ref{sec:artifacts}).

We implemented a~late-fusion strategy at the study level by taking an~unweighted average of the A4C model prediction and the PLAX ensemble prediction. This simple approach required no additional training and preserved the independence of the single-view models. 

As summarized in Table~\ref{tab:multiview_results}, the fusion outperformed both single-view baselines, reducing MAE to 6.37\% and increasing Pearson correlation to 0.709. This demonstrates that integrating complementary views yields a tangible gain in study-level EF prediction accuracy.

The fusion gain is incremental: the two views share overlapping information about LV systolic function, and the PLAX model is trained from A4C-derived proxy labels. Still, the improvement shows that simple study-level late fusion can add signal beyond either single-view prediction alone.

\begin{table}[htbp]
  \centering
    \caption{Study-level performance comparison of single-view and multi-view EF prediction. \emph{Evaluated on the 284-study multi-view subset (studies containing both A4C and PLAX); results are not directly comparable to Section~\ref{sec:plax-results}, which uses the full PLAX-only ground-truth cohort containing 295 studies.}}
    \label{tab:multiview_results}
  \begin{tabular}{l|c|c}
    \bfseries Model & \bfseries MAE & \bfseries Correlation \\
    \hline
    A4C-only & 7.01\% & 0.657 \\
    PLAX-only (ensemble) & 6.77\% & 0.676 \\
    \textbf{A4C + PLAX (fusion)} & \textbf{6.37\%} & \textbf{0.709} \\
  \end{tabular}
\end{table}

\begin{figure}[htbp]
\centerline{\includegraphics[width=\columnwidth]{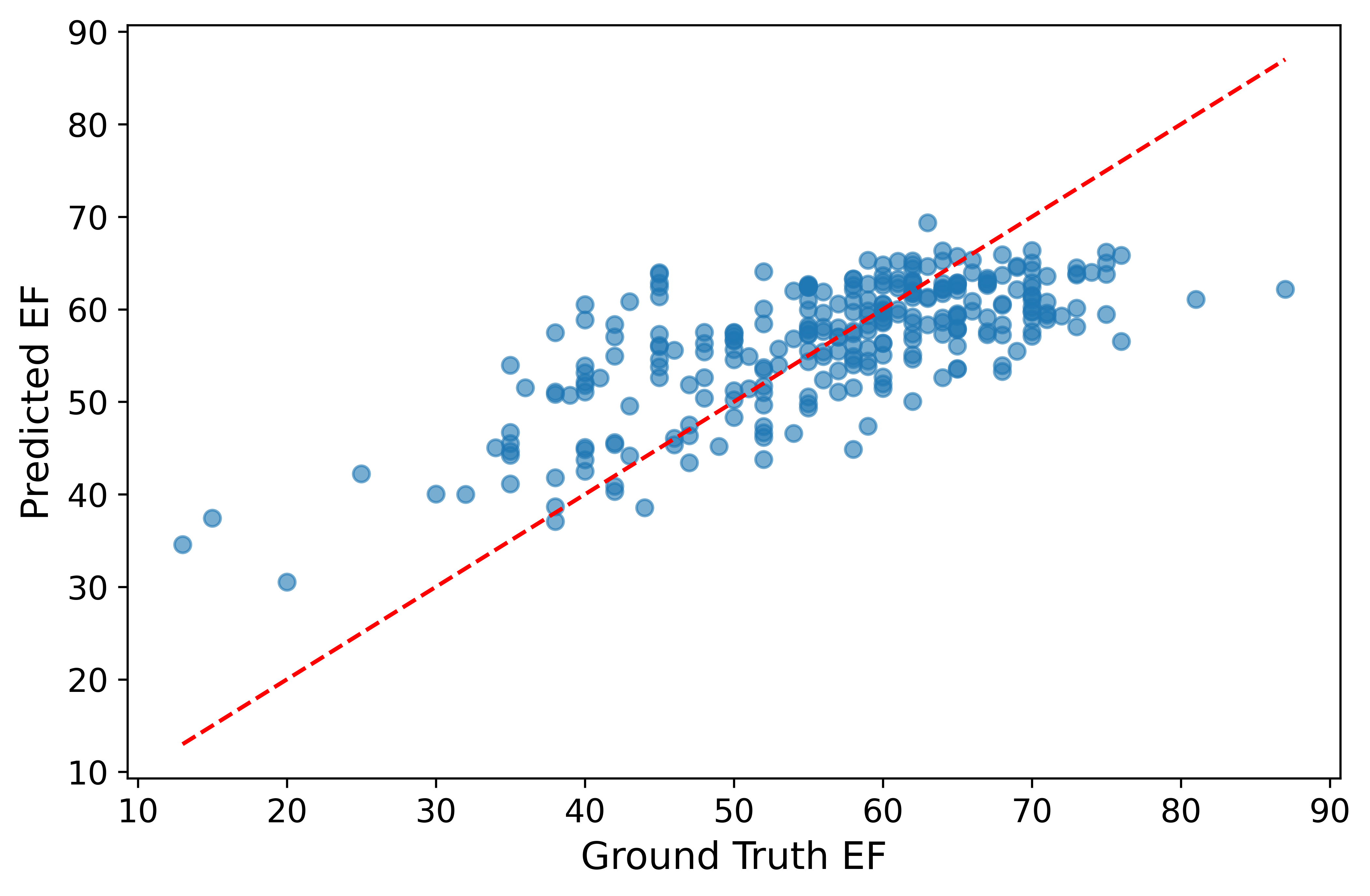}}
\caption{Scatter plot for A4C + PLAX fusion EF predictions against ground truth. The dashed red line indicates the line of identity.}
\label{fig9}
\end{figure}

\begin{figure}[htbp]
\centerline{\includegraphics[width=\columnwidth]{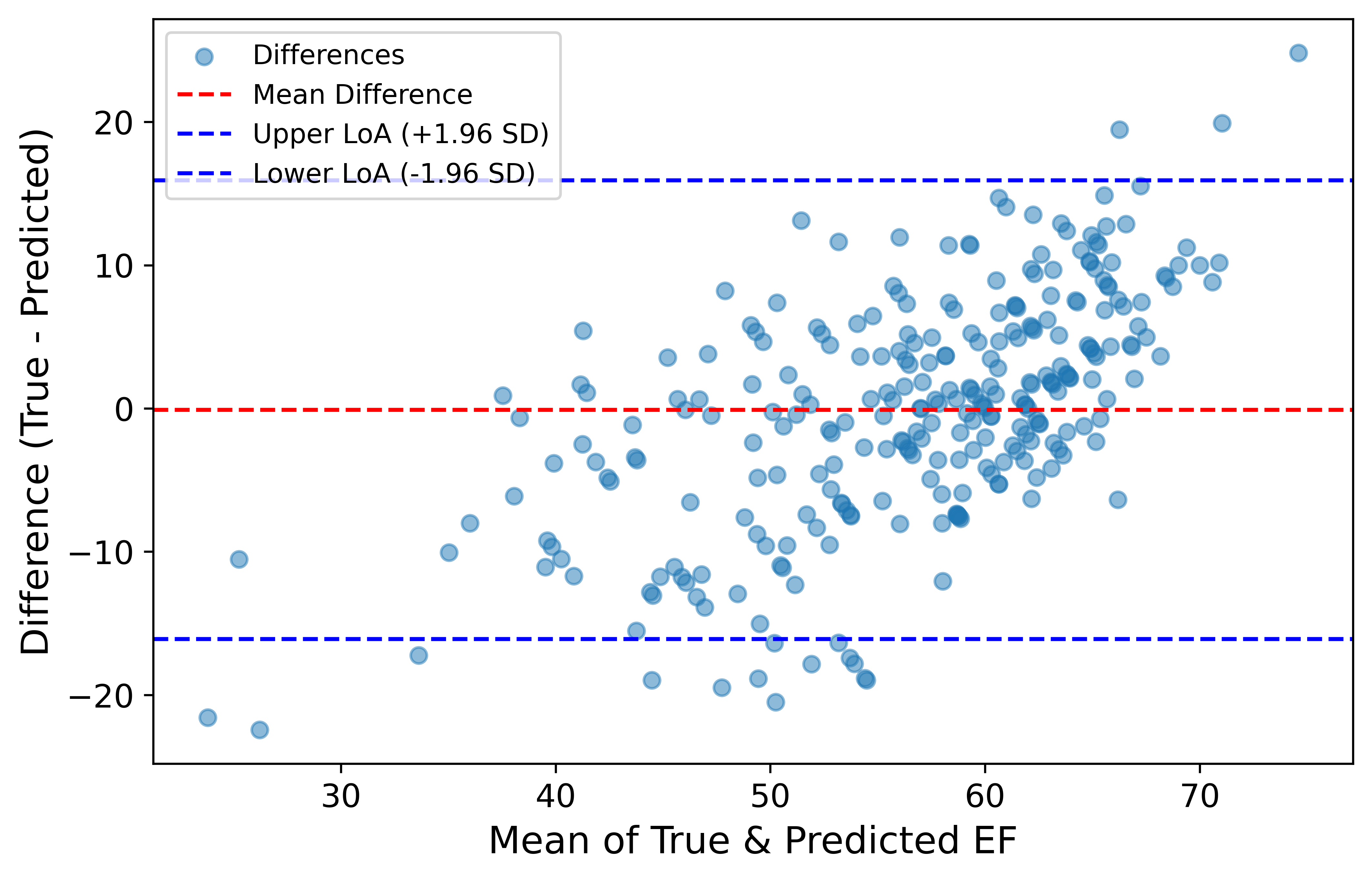}}
\caption{Bland-Altman plot for A4C + PLAX fusion EF predictions.}
\label{fig10}
\end{figure}

\paragraph{Agreement and calibration.}
The scatter plot (Fig.~\ref{fig9}) shows that fusion predictions align closely with the line of identity, with visibly reduced dispersion compared to single-view models. The Bland-Altman analysis (Fig.~\ref{fig10}) confirms this improvement: the mean bias was –0.08\%, with limits of agreement at +15.93\% and –16.08\%. Compared to the PLAX-only results, both the bias and error spread were reduced, indicating better calibration and more consistent predictions across the EF range.  
\paragraph{Single-view comparison.}
One thing worth noting is the comparison between the two single views. Although the PLAX-only ensemble reports a numerically lower MAE than the A4C-only model (6.77\% vs.\ 7.01\%) on this subset, this difference is not statistically significant: on the 284 paired studies, a study-level paired comparison of absolute errors yields no significant difference (Wilcoxon signed-rank $p=0.66$; paired $t$-test $p=0.48$; $\Delta$MAE $=0.24$\%, 95\% bootstrap CI $[-0.40, 0.88]$). The two single views are therefore statistically comparable rather than one being superior; in particular, the PLAX student matches but does not exceed its A4C teacher, consistent with proxy supervision introducing no view-level accuracy loss.

\paragraph{Proportional bias.}
The Bland--Altman plot (Fig.~\ref{fig10}) also indicates a proportional bias, with overestimation at lower EF values and underestimation at higher EF values. To quantify this, we regressed the per-study prediction differences (predicted minus reference EF) on the mean EF for each setting. All three views exhibit a statistically significant negative slope:
\begin{itemize}
    \item \textbf{A4C:} slope $-0.22$ (95\% CI $[-0.33,-0.12]$, $p<0.001$);
    \item \textbf{PLAX:} slope $-0.88$ (95\% CI $[-0.97,-0.80]$, $p<0.001$);
    \item \textbf{Fusion:} slope $-0.58$ (95\% CI $[-0.67,-0.49]$, $p<0.001$).
\end{itemize}
The pattern across views is informative as to cause. The A4C model, trained on expert-annotated EchoNet-Dynamic EF labels, still exhibits a significant slope, indicating that MSE-driven shrinkage toward the population mean is present even under clean supervision. The PLAX model, trained on A4C-derived proxy labels, shows the steepest slope, consistent with proxy supervision and note-derived label noise adding further shrinkage on top of this baseline; the fusion result falls between the two, as expected from averaging the two views.

\paragraph{Inter-view disagreement.}
As a further exploratory analysis, we tested whether the disagreement between the two single-view predictions, $|\mathrm{EF}_{\mathrm{A4C}} - \mathrm{EF}_{\mathrm{PLAX}}|$, is associated with the final fusion error. The association is weak and small in magnitude: the regression slope is $\approx 0.08$ EF points of error per EF point of disagreement (Pearson $r=0.07$, $p=0.21$; Spearman $r=0.13$, $p=0.04$), and a quartile analysis shows no consistently monotonic trend. Inter-view disagreement may therefore provide at most a weak reliability signal and is not, on this dataset, sufficient to serve as a standalone uncertainty measure.

\paragraph{Alternative fusion rules.}
We further examined whether a more elaborate fusion rule would improve on the unweighted average. On the 284 common studies, a learned convex weight $w,\mathrm{EF}*{\mathrm{A4C}} + (1-w),\mathrm{EF}*{\mathrm{PLAX}}$ selected $w \approx 0.44$, nearly recovering the equal-weight average, and did not reduce study-level MAE (5-fold cross-validated $6.38\%$ vs.\ $6.37\%$; Wilcoxon signed-rank $p=0.38$). A regularized linear fusion (ridge regression over both view predictions with an intercept) likewise showed no significant difference ($6.34\%$ vs.\ $6.37\%$; $p=0.87$). Learned fusion rules can in principle calibrate view-specific biases and assign data-driven weights, but they also introduce additional fitted parameters and greater overfitting risk in this limited paired-data setting. In contrast, the unweighted average is less flexible but parameter-free and preserves the independence of the single-view models.

\paragraph{Summary.}
Together, these analyses highlight the complementary value of integrating multiple echocardiographic views. The simple study-level fusion improved over both single-view baselines without additional video-model training. Beyond establishing a~benchmark for PLAX EF prediction, our results suggest that multi-view integration can serve as a practical strategy for more dependable EF assessment in real-world echocardiography workflows.

\FloatBarrier

\section{Artifacts and Reproducibility} \label{sec:artifacts}

To support reproducibility, transparency, and further research, we release a
complete set of artifacts that enable reconstruction of the evaluation cohorts via released label files (given authorized access to MIMIC-IV resources), and convenient inference with pretrained A4C and PLAX EF models. A concise usage guide is provided in Appendix~\ref{app:artifacts_usage}.

\begin{itemize}
    \item \textbf{GitHub repository.}
    We provide dataset label files, instructions for locating the corresponding videos within MIMIC-IV-Echo under the data-use agreement, and the GPT prompts used for EF extraction from clinical notes.
    
    \url{https://github.com/Jeffrey4899/PLAX_EF_Labels_202509}

    \item \textbf{Hugging Face Space.}
    A browser-based interactive demo for running inference with the released EF models (A4C/PLAX/multi-view fusion) on sample videos or user-uploaded inputs (research and education use only).
    
    \url{https://huggingface.co/spaces/Jeff4899/202509_PLAX_EF_Demo}

    \item \textbf{Trained models on Hugging Face.}
    Inference-ready models for both PLAX EF and A4C EF prediction are released
    for direct download and integration into existing pipelines.
    
    \url{https://huggingface.co/Jeff4899/202509_PLAX_EF} \\
    \url{https://huggingface.co/Jeff4899/202509_A4C_EF}

    \item \textbf{Google Colab demo.}
    An interactive notebook that enables users to evaluate the released models
    on sample echocardiography videos or their own uploads, with full access to
    and modification of the underlying code.
    
    \url{https://colab.research.google.com/drive/1E2IWrfpBIKI4cBoBTCn3OLwEK9o3GTMM?usp=sharing}
\end{itemize}

Together, these artifacts are intended not only to facilitate reproduction of the reported results, but also to serve as a foundation for
future methodological and clinical extensions.

\section{Conclusion}
We introduced, to the best of our knowledge, the first publicly available label resource enabling EF estimation from parasternal long-axis (PLAX) cine
echocardiography, together with a reproducible pipeline that learns from scarce labels by mining views, extracting EF from clinical notes, and training PLAX models with proxy supervision. On an independent note-derived ground-truth cohort, our PLAX ensemble achieved a study-level MAE of 6.86\%. We further showed that simple study-level late fusion of A4C and PLAX predictions improves accuracy to 6.37\% MAE (correlation 0.709) on studies containing both views, highlighting the practical value of multi-view integration for EF estimation.

We release label files, pretrained model weights, and runnable inference demos to facilitate reproduction of the reported evaluation (given authorized access to the underlying datasets) and to support downstream research. 

% --- insert in Conclusion, immediately before \paragraph{Limitations.} ---

\paragraph{Clinical utility.}
The clinical motivation for this work is based on two practical considerations. First, PLAX is a routinely acquired echocardiographic view, while EF estimation from apical views depends on sufficient apical image quality and can be affected by issues such as foreshortening \citep{lang2015ase}. Therefore, PLAX-based EF estimation may extend automated assessment to scans where standard apical windows are unavailable, incomplete, or suboptimal. Second, PLAX remains under-labeled for EF estimation in public resources. For example, EchoNet-LVH provides a large public PLAX video dataset, but its labels focus on chamber size and wall-thickness measurements rather than EF \citep{echonet-LVH}. To our knowledge, no public PLAX--EF benchmark currently exists. Therefore, this work provides a reproducible PLAX--EF benchmark, proxy-supervision pipeline, and baseline models for a clinically relevant but under-labeled view.

\paragraph{Limitations.}
Our PLAX models are trained with proxy supervision derived from an A4C teacher,
so performance is inherently affected by teacher noise and may be bounded by
the A4C predictor under domain shift. In addition, view classification errors
can propagate into both cohort construction and proxy labeling. Our ground-truth
evaluation relies on EF values extracted from clinical notes with a strict
time-window heuristic, which may still introduce residual label noise when EF
reporting and imaging are not perfectly aligned. All released artifacts are
intended for research and education use only; further external validation on
independent clinical datasets would be required before considering any clinical
deployment.

\paragraph{Future work.}
To further improve and validate our model, we are initiating a collaboration with a leading heart hospital to leverage their clinical datasets for refining label accuracy and enhancing generalizability. Future work will focus on incorporating expert-annotated PLAX EF values, for which a dataset is currently being secured. Additionally, we plan to evaluate clinical applicability through external validation and potential clinical trials, ensuring real-world effectiveness in echocardiographic workflows. 

Taken together, our results suggest that multi-view integration offers a practical and effective way to improve upon A4C-only prediction, while also demonstrating that PLAX views alone can provide meaningful EF estimates when A4C views are unavailable or suboptimal.

% Mandatory Sections. Please complete, especially for final publication
\acks{
    This work was supported by the Center for Sensing to Intelligence (S2I) at
    Caltech and by Charles R. Trimble. 
    
    The authors thank the contributors of the
    MIMIC-IV-Echo, EchoNet, and TMED-2 datasets for making their data publicly
    available. We also thank the MIDL 2025 reviewers for their constructive
    feedback, which helped improve the clarity and scope of this extended version.
}

\ethics{
This study used only publicly available, de-identified datasets
(MIMIC-IV-Echo, MIMIC-IV-Note, and EchoNet) released under their respective
data use agreements. Institutional review board approval was not required
because no new data were collected and no human subjects were directly
involved in this study. All analyses were conducted in compliance with the
ethical standards of the data providers and applicable regulations.
}

\coi{
The authors declare that they have no conflicts of interest.
}

\data{
All artifacts required to reproduce the results of this study (given authorized access to MIMIC-IV resources) are publicly available at the following URLs:

\begin{itemize}
    \item \textbf{GitHub repository (labels, prompts, instructions):} \\
    \url{https://github.com/Jeffrey4899/PLAX_EF_Labels_202509}

    \item \textbf{Hugging Face Space (interactive demo):} \\
    \url{https://huggingface.co/spaces/Jeff4899/202509_PLAX_EF_Demo}

    \item \textbf{Hugging Face models (PLAX EF):} \\
    \url{https://huggingface.co/Jeff4899/202509_PLAX_EF}

    \item \textbf{Hugging Face models (A4C EF):} \\
    \url{https://huggingface.co/Jeff4899/202509_A4C_EF}

    \item \textbf{Google Colab demo:} \\
    \url{https://colab.research.google.com/drive/1E2IWrfpBIKI4cBoBTCn3OLwEK9o3GTMM?usp=sharing}
\end{itemize}

Due to data use agreements, raw MIMIC-IV-Echo videos are not redistributed. The released label files provide identifiers and instructions for locating the corresponding samples under authorized access.
}

\bibliography{sample}

% =========================
% Appendix
% =========================
\clearpage
\appendix

\section{Training Details}\label{app:training_details}
This appendix documents implementation details and hyperparameters for all
trainable components in our pipeline. The goal is to enable faithful
reproduction while keeping the main text focused on methodology and results.

\subsection{Video View Classifier Training Details}\label{app:view_classifier_training}
This subsection summarizes training and implementation details for the
view-classification components used to mine A4C and PLAX clips from MIMIC-IV-Echo
(Section~\ref{VideoViewClassifier}). We focus on model training and data
preprocessing details; the manual reviewer verification results are reported in
the main text.

\subsubsection{TMED-2 Image View Classifier (ResNet-34)}\label{app:tmed2_image_classifier}

\paragraph{Task and labels.}
We first train an image-based view classifier on TMED-2 \citep{huangTMED2Dataset2022}
to bootstrap high-confidence view candidates from unlabeled MIMIC-IV-Echo videos.
The classifier predicts five categories: A4C, A2C, PLAX, PSAX, and a combined
``A4C/A2C/OTHER'' class. In our implementation, string labels are mapped to class
indices as follows: \texttt{A4C}$\rightarrow 0$, \texttt{A2C}$\rightarrow 1$,
\texttt{PLAX}$\rightarrow 2$, \texttt{PSAX}$\rightarrow 3$, and
\texttt{A4CorA2CorOther}$\rightarrow 4$.

\paragraph{Model and architectural adjustments.}
We use a ResNet-34 backbone initialized with ImageNet weights. To better match
the relatively small echocardiography frame resolution and avoid overly
aggressive early downsampling, we replace the first convolution with a
$3{\times}3$ kernel (stride 1, padding 1) and remove the initial max-pooling
layer. The final fully connected layer is replaced with a linear classifier
outputting 5 logits.

\paragraph{Preprocessing.}
Each frame is resized to $112{\times}112$. Because TMED-2 images may be stored as
single-channel intensity images, we convert inputs to 3 channels (grayscale
replication) and apply standard ImageNet normalization after converting to a
tensor.

\paragraph{Label smoothing and loss.}
To improve robustness to label noise and the heterogeneous ``A4C/A2C/OTHER''
category, we apply label smoothing with $\epsilon=0.05$. Specifically, each hard
label is converted into a 5-dimensional target distribution where the ground-truth
class receives probability $1-\epsilon$ and the remaining mass $\epsilon$ is
distributed uniformly over the other classes. The model output logits are passed
through \texttt{log\_softmax}, and we optimize the KL divergence between predicted
log-probabilities and the smoothed target distribution using KLDivLoss with
\texttt{reduction=batchmean}.

\paragraph{Optimization and training protocol.}
We follow the official TMED-2 split indicators provided in our per-image CSV
metadata, training on the \texttt{train} subset and monitoring loss on the
\texttt{val} subset. Optimization uses SGD with momentum 0.9, initial learning
rate $10^{-3}$, batch size 32, and 50 epochs. We apply a StepLR schedule with
step size 10 and decay factor 0.1. The checkpoint with the lowest validation
loss is saved as the final image model.

\paragraph{Implementation notes and compute.}
Training is implemented in PyTorch with \texttt{DataParallel}. In our setup, the
training runs on one NVIDIA A100 GPUs. Data loading uses 1 worker process.
Validation curves and learning-rate schedules are logged via Weights \& Biases
for traceability.

\subsubsection{Video View Classifier (X3D-s)}\label{app:x3ds_video_classifier}

\paragraph{Training data.}
We train a three-way video classifier for \texttt{A4C}/\texttt{PLAX}/\texttt{OTHER}
using the multi-dataset strategy described in Section~\ref{VideoViewClassifier}:
A4C videos from EchoNet-Dynamic \citep{echonet-dynamic}, PLAX videos from
EchoNet-LVH \citep{echonet-LVH}, and \texttt{OTHER} videos mined from MIMIC-IV-Echo
using the TMED-2 image classifier with conservative score thresholds to reduce
contamination.

\paragraph{Clip sampling and preprocessing.}
Videos are decoded into RGB frames and converted into fixed-length clips with
$L{=}64$ frames using temporal stride 2. For sufficiently long videos, a random
start frame is sampled to provide temporal augmentation; if the decoded clip is
shorter than $L$, missing frames are padded with black frames to ensure a fixed
temporal input length. Frame-level transforms (e.g., padding/cropping and
normalization) are applied independently to each frame and then stacked into a
$L{\times}C{\times}H{\times}W$ tensor before being permuted to the
$C{\times}L{\times}H{\times}W$ format expected by 3D backbones.

\paragraph{Model and training.}
We fine-tune a pretrained X3D-s network \citep{feichtenhofer2020x3dexpandingarchitecturesefficient}
by replacing the final projection layer with a three-logit classifier head for
\texttt{A4C}/\texttt{PLAX}/\texttt{OTHER}. The training objective is standard
multi-class classification, and we select the final checkpoint based on
validation performance on a held-out split.

\paragraph{Implementation notes and compute.}
Video-model training is implemented in PyTorch and run on three NVIDIA A100 GPUs
using data-parallel training. Data loading uses 8 worker processes. Training
progress and validation metrics are logged via Weights \& Biases.

\subsection{A4C EF Regressor (EchoNet-Dynamic)}\label{app:train_a4c}

\paragraph{Backbone and regression head.}
We use the TorchVision \texttt{r2plus1d\_18} R(2+1)D architecture initialized
from a public video-pretrained checkpoint. The final classification layer is
replaced with a single-neuron linear layer to output a scalar EF prediction.

\paragraph{Clip sampling and preprocessing.}
Given an input video, we decode RGB frames and construct a fixed-length clip of
$L{=}64$ frames using temporal stride 2 (i.e., sampling every other frame from a
128-frame window). For sufficiently long videos, the start frame is randomly
sampled to provide temporal augmentation; for shorter clips, missing frames are
padded with black frames to maintain a consistent input length. Each frame is
converted to channel-first format and normalized after scaling to $[0,1]$.
Spatial augmentation follows a lightweight padding-and-cropping strategy at a
target resolution of $112{\times}112$: with equal probability, we either
(i) pad by 24 pixels and crop a fixed top-right $112{\times}112$ region, or
(ii) pad by 12 pixels and apply a random $112{\times}112$ crop. During
evaluation, we disable spatial augmentation and use only normalization.

\paragraph{Optimization and model selection.}
We train on the official EchoNet-Dynamic \texttt{TRAIN} split, monitor
performance on \texttt{VAL}, and report final performance on \texttt{TEST}
(75\%/12.5\%/12.5\%). The training objective is mean squared error (MSE) between
predicted and ground-truth EF values (percentage points). We optimize with RAdam
(initial learning rate $10^{-3}$, batch size 32) for 100 epochs. A
ReduceLROnPlateau scheduler (patience 5 epochs, reduction factor 0.1) adapts the
learning rate based on validation loss. The checkpoint with the lowest
validation loss is selected as the final A4C model.

\paragraph{Implementation notes.}
All models are implemented in PyTorch and trained on a single NVIDIA H100 GPU.
Data loading uses 8 worker processes. Training progress, including validation
metrics and learning-rate schedules, is logged using Weights \& Biases to ensure traceability and experiment reproducibility.

\subsection{PLAX EF Model Training Details}\label{app:plax_training_details}
This subsection documents implementation details for the PLAX EF regression
models reported in Section~\ref{sec:plax-results}. We train two models with
identical architectures and training protocols, differing only in batch size
(16 vs.\ 32); the final PLAX predictor is formed by simple unweighted ensembling.

\paragraph{Training data and proxy targets.}
PLAX training clips are constructed as described in Section~\ref{2.5}. Each PLAX
video inherits a \emph{study-level} proxy EF label computed by averaging the
A4C teacher model predictions across all A4C clips available in the same study.
The resulting PLAX training set is split into \texttt{TRAIN}/\texttt{VAL} at the
study level (Section~\ref{2.5}) so that multiple clips from the same study do
not leak across splits.

\paragraph{Architecture.}
Both PLAX models use the TorchVision \texttt{r2plus1d\_18} R(2+1)D backbone
initialized from a public video-pretrained checkpoint. We replace the final
classification layer with a single-neuron linear head to output a scalar EF
prediction.

\paragraph{Clip sampling and preprocessing.}
Given an input PLAX video, we decode RGB frames and construct a fixed-length
clip of $L{=}64$ frames using temporal stride 2. For sufficiently long videos,
the start frame is randomly sampled to provide temporal augmentation; for
shorter videos, missing frames are padded with black frames to maintain a
consistent temporal input length. Frame-wise preprocessing is applied before
stacking frames into a clip tensor. In our implementation, we use padding
followed by random cropping to a fixed spatial resolution, together with
intensity scaling to $[0,1]$ and channel-wise normalization.

\paragraph{Optimization and model selection.}
We train both models using mean squared error (MSE) loss between predicted EF
and proxy EF labels. Optimization uses the RAdam optimizer with initial learning
rate $10^{-3}$ and a maximum of 100 epochs. We apply a ReduceLROnPlateau learning
rate scheduler (patience 5 epochs, reduction factor 0.1) based on validation
loss. The checkpoint with the lowest \texttt{VAL} loss is selected for each
configuration. The two configurations reported in Table~\ref{tab:model_training}
differ only by batch size (16 vs.\ 32); all other settings are kept fixed.

\paragraph{Study-level inference and ensembling.}
At inference time, we compute study-level predictions by averaging video-level
EF predictions across all PLAX clips within each study. The final PLAX estimate
is then obtained by an unweighted average of the two study-level predictions
from the batch-size-16 and batch-size-32 models (i.e., a two-model ensemble).

\paragraph{Implementation notes and compute.}
Training is implemented in PyTorch. The PLAX models are trained on a single
NVIDIA H100 GPU, with data loading using 8 worker processes. Training curves and
validation metrics are logged via Weights \& Biases for traceability.

% =========================
% Appendix B: Artifacts
% =========================
\section{Artifact Access and Usage}\label{app:artifacts_usage}
This appendix summarizes what is released and how to use the artifacts listed
in Section~\ref{sec:artifacts}. The released resources primarily support
\emph{inference and evaluation} with pretrained models and released label files.
We do not redistribute raw MIMIC-IV-Echo videos due to the PhysioNet data-use
agreements. An overview of the released artifacts is provided in
Table~\ref{tab:artifact_overview}, and a screenshot of the Hugging Face Space
demo interface is shown in Fig.~\ref{fig:hf_space_demo}.

\subsection{Overview of released artifacts}\label{app:artifact_overview_subsec}
Table~\ref{tab:artifact_overview} provides a concise overview of the released
artifacts. URLs are listed in Section~\ref{sec:artifacts} to avoid duplication.

\begin{table*}[t]
\centering
\caption{Overview of released artifacts (URLs are listed in Section~\ref{sec:artifacts}).}
\label{tab:artifact_overview}
\begin{tabular}{p{0.20\textwidth} p{0.75\textwidth}}
\toprule
\textbf{Artifact} & \textbf{Summary} \\
\midrule
GitHub repository &
CSV label files for reconstructing PLAX/A4C cohorts in MIMIC-IV-Echo (proxy-labeled PLAX train/val, note-derived ground-truth cohorts, and the paired multi-view subset), plus EF-extraction prompts/notebooks and instructions for locating videos under authorized access. \\[2pt]

Hugging Face model weights &
Pretrained checkpoints for A4C EF inference and PLAX EF inference. The main paper uses a two-checkpoint PLAX ensemble (batch sizes 16 and 32) and averages their study-level predictions. \\[2pt]

Hugging Face Space demo &
Browser-based UI for uploading short PLAX and/or A4C clips and obtaining per-view EF predictions, as well as their mean when both views are provided (research and education use only). \\[2pt]

Google Colab notebook &
Runnable notebook that downloads released weights and demonstrates inference on sample clips or user-provided uploads; can be adapted to compute metrics given authorized access to the underlying datasets. \\
\bottomrule
\end{tabular}
\end{table*}

\subsection{Hugging Face Space demo interface}\label{app:hf_space_interface}
\begin{figure*}[t]
\centering
\includegraphics[width=\textwidth]{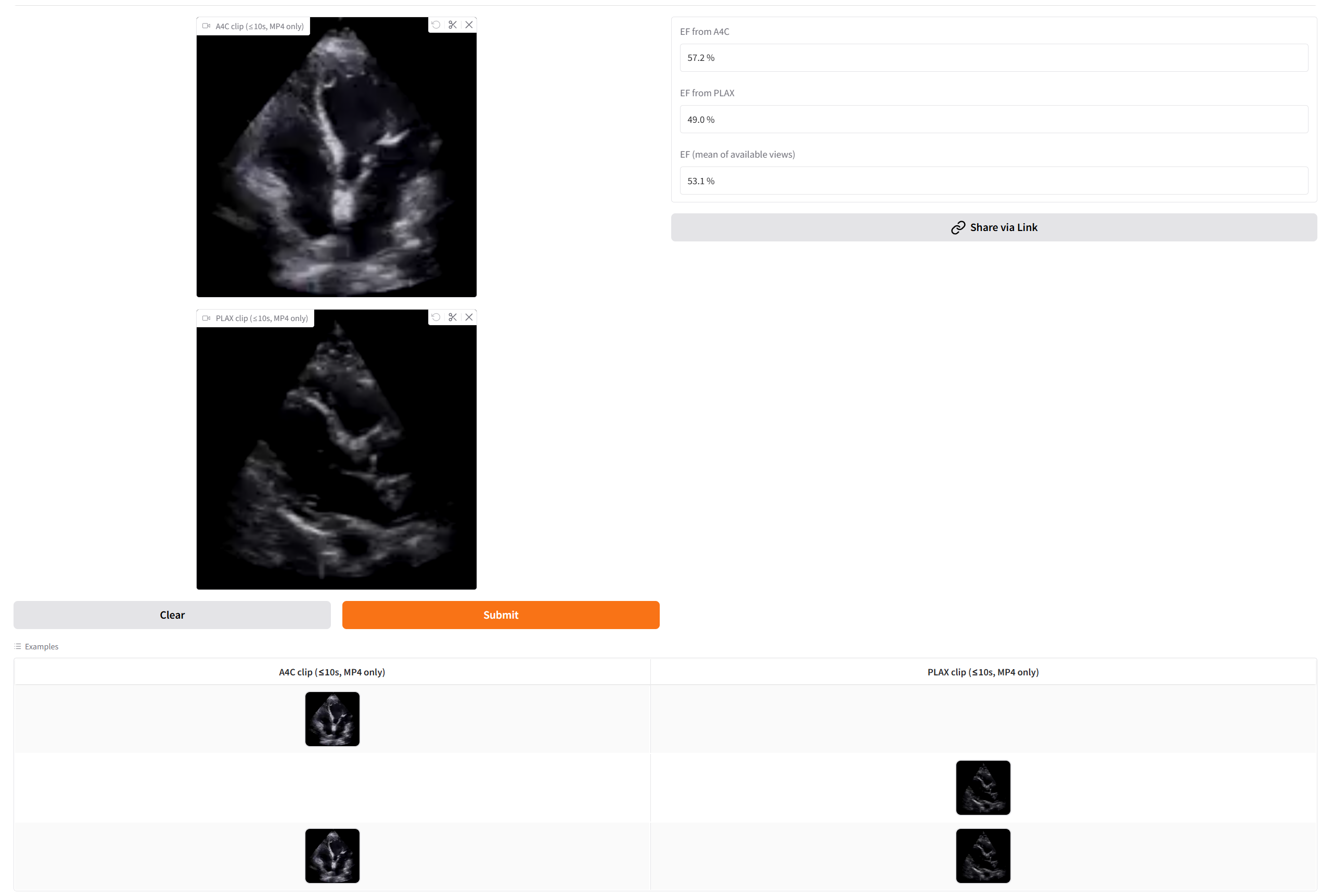}
\caption{Screenshot of the Hugging Face Space demo for EF estimation from PLAX and A4C clips. The demo reports \texttt{EF from A4C}, \texttt{EF from PLAX}, and \texttt{EF (mean of available views)} when both are provided. The demo is intended for research and education use only and is not a medical device.}
\label{fig:hf_space_demo}
\end{figure*}

Fig.~\ref{fig:hf_space_demo} shows the Hugging Face Space interface used for
online inference. Users may upload a PLAX clip and/or an A4C clip and obtain
\texttt{EF from A4C}, \texttt{EF from PLAX}, and \texttt{EF (mean of available views)}.
The Space is intended for research and education use only and is not a medical
device.

\subsection{Reproducing the reported evaluation}\label{app:reproduce_eval}
To reproduce the metrics reported in Section~\ref{sec:results}, the required steps are:
\begin{itemize}
    \item \textbf{Obtain authorized access.} Obtain credentialed access to
    MIMIC-IV-Echo and (if needed) MIMIC-IV-Note under the PhysioNet data-use agreements.
    \item \textbf{Reconstruct cohorts from label files.} Use the released CSV
    label files to locate the corresponding videos in MIMIC-IV-Echo (e.g., via
    \texttt{subject\_group}, \texttt{subject\_id}, \texttt{study\_id}, and \texttt{file\_id},
    depending on the label format).
    \item \textbf{Run inference with released weights.} Download the pretrained
    A4C and PLAX model checkpoints and run inference on the target clips.
    \item \textbf{Compute study-level metrics.} Aggregate video-level predictions
    by averaging within each study (as in Section~\ref{sec:results}) and report MAE
    and correlation on the note-derived ground-truth cohort.
\end{itemize}

\subsection{Running inference without local setup}\label{app:run_inference_no_setup}
\paragraph{Hugging Face Space.}
Open the Space URL (Section~\ref{sec:artifacts}), upload a PLAX clip and/or an A4C
clip (short MP4 recommended), and click \texttt{Submit}. The demo reports
\texttt{EF from A4C}, \texttt{EF from PLAX}, and the mean when both are available.

\paragraph{Google Colab.}
Open the Colab notebook URL (Section~\ref{sec:artifacts}) and run all cells. The
notebook downloads the released weights and runs inference on sample inputs or
user-provided clips. For simplicity, lightweight demos may use one PLAX model and
one A4C model, whereas the main paper results use the full aggregation setup
(two PLAX checkpoints plus the A4C model, as described in the paper).

\subsection{Practical notes}\label{app:practical_notes}
\begin{itemize}
    \item \textbf{Input clip length and format.} The Hugging Face demo is intended
    for short clips (e.g., $\leq 10$s) in MP4 format. Very long clips may time out
    or increase latency in online inference environments.
    \item \textbf{View requirements.} The released models are trained for A4C and PLAX
    views. If the uploaded clip is not a true A4C/PLAX view (or is a partial/atypical
    acquisition), predictions may be unreliable. For rigorous evaluation, we
    recommend using view-filtered cohorts and study-level aggregation as described
    in Section~\ref{sec:results}.
    \item \textbf{Meaning of demo outputs.}
    \texttt{EF from A4C} and \texttt{EF from PLAX} are per-view EF predictions produced by
    the corresponding view-specific models. \texttt{EF (mean of available views)}
    is a simple unweighted average of the available view predictions (if only one
    view is provided, it equals that view’s estimate).
    \item \textbf{Study-level aggregation.} The main paper reports \emph{study-level}
    results by averaging predictions over multiple clips within the same study.
    Single-clip predictions (e.g., via the demo) are best interpreted as qualitative
    examples rather than a substitute for study-level reporting.
    \item \textbf{Privacy and intended use.} All artifacts are intended for research
    and education use only and must not be used for clinical diagnosis or treatment.
    Users are responsible for ensuring that any uploaded data (e.g., to online demos)
    comply with applicable privacy policies, regulations, and data-use agreements.
    When in doubt, prefer running inference locally or in Colab on appropriately
    de-identified data.
\end{itemize}

\subsection{Notes on data redistribution and intended use}\label{app:data_redistribution_notes}
Raw MIMIC-IV-Echo videos are not redistributed due to data-use agreements. The
released label files provide identifiers and instructions for locating the corresponding
samples under authorized access. All released artifacts are intended for research and
education use only and are not medical devices.

\end{document}